\documentclass[runningheads]{llncs}

\usepackage{amssymb}
\input{parsetree.sty}

\newcommand{\npile}[1]{%
  \ptnode{               %
    \( \begin{array}{c}#1%
       \end{array}       %
    \)   }}

\newcommand{\ignore}[1]{}
\newcommand{\nit}[1]{{\it #1}}
\newcommand{\boxtheorem}{\hfill $\Box$}
\newcommand{\po}{\hspace{-2mm}{\bf .}~ }

\newcommand{\bi}{\begin{itemize}}
\newcommand{\ei}{\end{itemize}}
\newcommand{\be}{\begin{enumerate}}
\newcommand{\ee}{\end{enumerate}}
\newcommand{\bax}{\begin{ax}}
\newcommand{\eax}{\end{ax}}

\newcommand{\Xstar}[1]{\mbox{$K^\star$}}

\newcommand{\comlb}[1]{{\vspace{4mm}\noindent \bf COMM(LEO):}~ {\em #1 }\hfill {\bf END.}\\}
\newcommand{\comcs}[1]{{\vspace{4mm}\noindent \bf COMM(CAMILLA):}~ {\em #1 }\hfill {\bf END(CAMILLA.}\\}
\newcommand{\IC}{{\it IC}}
\newcommand{\opb}{\nit{op}(B)}
\newcommand{\TP}{{\it TP}}

\newtheorem{lem}{Lemma}

\newtheorem{ax}{Axiom}

\newtheorem{exa}{Example }

 \title{Database Repairs and  Analytic Tableaux}

\author{{\bf Leopoldo Bertossi}\inst{1}
\and {\bf Camilla Schwind}\inst{2}}
\authorrunning{Bertossi and Schwind}

\institute{Carleton University\\ School of Computer Science\\
Ottawa, Canada K1S 5B6\\ bertossi@scs.carleton.ca \vspace{2mm}
\and MAP-CNRS, Ecole d'architecture de Marseille\\ 183, Avenue de
Luminy, \\ 13288 Marseille, Cedex 9, France\\
schwind@map.archi.fr}
\date{}

\begin{document}

\maketitle

\ignore{
\begin{opening}

\title{Database Repairs and Analytic Tableaux}

\author{\surname{Leopoldo Bertossi} \email{bertossi@scs.carleton.ca}}
\institute{Carleton University, School of Computer Science,
Ottawa, Canada}
\author{\surname{Camilla Schwind} \email{schwind@map.archi.fr}}
\institute{MAP-CNRS, Ecole d'architecture de Marseille, 183,
Avenue de Luminy, 13288 Marseille, Cedex 9, France}

\runningtitle{Database Repairs} \runningauthor{Bertossi \&
Schwind} }

\begin{abstract}
In this article, we characterize in terms of analytic tableaux the
repairs of inconsistent relational databases, that is databases
that do not satisfy a given set of integrity constraints. For this
purpose we provide closing and opening criteria for branches in
tableaux that are built for database instances and their integrity
constraints. We use the tableaux based characterization as a basis
for  consistent query answering, that is for
retrieving from the database answers to queries that are
consistent wrt the integrity constraints.
\end{abstract}

%\end{opening}

\section{Introduction}\label{sec:intro}

 The notion of consistent answer to a query posed to
an inconsistent database was defined in
\cite{ArenasBertossiChomicki:99}: A tuple is a consistent answer
if it is an answer, in the usual sense, in every possible repair
of the inconsistent database. A repair is a new database instance
that satisfies the integrity constraints and differs from the
original instance by a minimal set of changes wrt set inclusion.

A computational methodology to obtain such consistent answers was
also presented in \cite{ArenasBertossiChomicki:99}. Nevertheless,
it has some limitations in terms of the syntactical form of
integrity constraints and queries it can handle. In particular, it
does not cover the case of existential queries and constraints.

In classical logic, analytic tableaux \cite{Beth:59} are used as a
formal deductive system for propositional and predicate logic.
Similar in spirit to resolution, but with some important
methodological and practical differences \cite{FittingJAR:88},
they are mainly used for producing formal refutations from a
contradictory set of formulas. Starting from a set of formulas,
the system produces a tree with formulas in its nodes. The set of
formulas is inconsistent whenever all the branches in the tableau
can be closed. A branch closes when it contains a formula and its
negation.

In this paper we extend the tableaux methodology to deal with a
relational database instance plus a set of integrity constraints
that the first fails to satisfy. Consequently, both inputs
together can be considered as building an inconsistent set of
sentences. In this situation, we give criteria for closing
branches in a tableau for a relational database instance.

The technique of ``opening tableaux" was  introduced in
\cite{LafonSchwind:88} for a solution to the frame  problem, and
in \cite{Schwind:90,RischSchwind:94} for applying tableaux methods
to default logic. In this paper we show how to open tableaux for
database instances plus their constraints, and this notion of
opening is applied to characterize and represent by means of a
tree structure all the repairs of the original database. Finally,
we sketch how this representation could be used to retrieve
consistent query answers. At least at the theoretical level, the
methodology introduced in this paper could be applied to any kind
of first order (FO) queries and constraints.

This paper is organized as follows. In  section
\ref{se:InconsistentDatabasesandRepairs}, we define our notion of
repair of a inconsistent  database. Section
\ref{se:DatabaseRepairsandAnalyticTableaux} recalls the definition
of analytic tableaux and shows how databases and their repairs can
be characterized as openings of closed tableaux. In section
\ref{se:Complexity} we show the relationship between consistent
query answering and Winslett's approach to knowledge base update;
this allows us to obtain some complexity results for our
methodology. Section \ref{se:QueryAnswering} shows how consistent
answers to queries posed to an inconsistent database can be
obtained using the analytic tableaux.  In section \ref{sec:circum}
we show the relationship of  consistent query answering with
minimal entailment, more specifically, in section
\ref{sec:circsp}, with circumscriptive reasoning. This yields a
method for implementing the approach, which is studied in  section
\ref{sec:implementation}.

\section{Inconsistent Databases and Repairs}\label{se:InconsistentDatabasesandRepairs}

In this paper  a database instance is given by a finite set of
finite relations on a database schema. A database schema can be
represented in logic by a typed first-order language, ${\cal L}$,
containing a finite set of sorted database predicates and a fixed infinite
set of constants $D$. The language contains a predicate for each
database relation and the constants in $D$ correspond to the
elements in the database domain, that will be also denoted by $D$.
That is every database instance has an infinite domain $D$. We
also have a set of integrity constraints $\IC$ expressed in
language ${\cal L}$. These are first-order formulas which the
database instances are expected to satisfy. In spite of this,
there are realistic situations where a database may not satisfy
its integrity constraints \cite{ArenasBertossiChomicki:99}. If a
database instance satisfies $\IC$, we say that it is consistent
(wrt $\IC$), otherwise we say it is inconsistent. In any case, we
will assume from now on that $\IC$ is a consistent set of first
order sentences.

A database instance $r$ can be represented by a finite set of
ground atoms in the database language, or alternatively, as a
Herbrand structure over this language, with Herbrand domain $D$
\cite{lloyd87}.
In consequence, we can say
that a database instance $r$ is consistent, wrt $\IC$, when its
corresponding Herbrand structure is a model of $\IC$, and we write
$r \models \IC$.

The active domain of a database instance $r$ is the set of those
elements of $D$ that explicitly appear (in the extensions of the database
predicates) in $r$. The active domain
is always finite and we denote it by ${\it Act}(r)$. We may also
have a set of built-in (or evaluable) predicates, like equality,
arithmetical relations, etc.  In this case, we have the language
${\cal L}$ possibly extended with these predicates. In all
database instances each of these predicates has a fixed  and
possibly infinite extension. Of course, since we defined database
instances as finite sets of ground atoms, we are not considering
these built-in atoms as members of database instances.

In database applications, it is usually the case that an
inconsistent database\footnote{Sometimes we will simply say
``database" instead of ``database instance".} has ``most" of its
data contents still consistent wrt $\IC$ and can still provide
``consistent answers" to queries posed to it. The notion of
consistent answer was defined and analyzed in
\cite{ArenasBertossiChomicki:99}. This was done on the basis of
considering all possible changes to $r$, in such a way that it
becomes a consistent database instance. A consistent answer is an
answer that can be retrieved from all those repairs that differ
from the original instance in a minimal way.

The notion of minimal change, defined in
\cite{ArenasBertossiChomicki:99}, is  based on the notion of
minimal distance between models using symmetric set difference
~$\Delta$ of sets of database tuples.

\begin{definition}\label{de:close} \em \cite{ArenasBertossiChomicki:99}
Given databases instances\footnote{We are assuming here and
everywhere in the paper that all database instances have the same
predicates and domain.}
 $r$, $r'$ and $r''$, we say that $r'$ ~is closer to $r$
than $r''$ iff $r\Delta r'  ~\subseteq~ r \Delta r''$. This is
denoted by $r'{\leq}_{r} r''$. \boxtheorem
\end{definition}

It is easy to see that ${\leq}_{r}$ is an order relation. Only
database predicates are taken into account for the notion of
distance. This is because built-in predicates are not subject to
change; and then they have the same extension in all database
instances. Now we can define the ``repairs" of an inconsistent
database instance.

\begin{definition}\label{de:cons}
\em \cite{ArenasBertossiChomicki:99}\\ (a) Given database
instances $r$ and $r'$, ~$r'$ is a {\em repair} of ~$r$, if ~$r'
\models IC$ and $r'$ is a  minimal element in the set of instances
wrt the order  ${\leq}_{r}$.\\ (b) Given a database instance $r$,
a set $IC$ and a first order query $Q(\bar{x})$, we say that a
ground tuple $\bar{t}$ is a consistent answer to $Q$ in $r$ wrt
$IC$ ~iff~ $r' \models Q[\bar{t}]$ for every repair $r'$ of $r$
(wrt $IC$). \boxtheorem
\end{definition}

\begin{example} \label{ex:repairs}
Consider the integrity constraint $$IC:~ \forall x, y, z
(Supply(x,y,z) ~\wedge~ Class(z,T_4) ~\rightarrow~ x=C),$$ stating
that $C$ is the only provider of items of class $T_4$; and the
inconsistent database $r  =  \{Supply(C, D_1,It_1), Supply(D, D_2,
It_2),$ $Class(It_1, T_4), Class(It_2,$ $T_4)\}$. We have only two
possible (minimal) repairs of the original database instance,
namely ~$r_1 = \{Supply(C, D_1,It_1), Class(It_1, T_4),
Class(It_2, T_4)\}$ ~and ~ $r_2 =$ $\{Supply(C, D_1,It_1),
Supply(D, D_2, It_2),$ $Class(It_1, T_4)\}$.

Given the query ~$Q(x,y,z): ~Supply(x,y,z)?$, the tuple $(C,
D_1,It_1)$ is a consistent answer because it can be obtained from
every repair, but $(D, D_2, It_2)$ is not, because it cannot be
retrieved from $r_1$. \boxtheorem
\end{example}

It is possible to prove \cite{ArenasBertossiChomicki:99} that for
every database instance $r$ and set $IC$ of integrity constraints,
there is always a repair $r'$. If $r$ is already consistent, then
$r$ is the only repair. The following lemma, easy to prove, will
be useful.

\begin{lem} \po \label{le:disord}\em
\begin{enumerate}
  \item If ~$r'{\le}_{r} r''$, then   ~$r\cap r'' \subseteq r\cap
  r'$.
  \item If ~$r' \subseteq r$, then ~$r \Delta r' = r \setminus r'$.
\boxtheorem
\end{enumerate}
\end{lem}

We have given  a semantic definition of consistent answer to a
query in an inconsistent database. We would like to compute
consistent answers, but via computing all possible repairs and
checking answers in common in {\bf all} of them. Actually there
may be an exponential number of repairs in the size of the
database \cite{scalar}.

In \cite{ArenasBertossiChomicki:99,CelleBertossi:2K} a mechanism
for computing and checking consistent query answers was
considered. It does not produce/use the repairs, but it queries
the only explicitly available inconsistent database instance.
Given a FO query $Q$, to obtain the consistent answers wrt  a finite
set of FO ICs $\IC$, $Q$ is
qualified with appropriate information derived from the
interaction between $Q$ and $\IC$. More precisely, if we want
the consistent answers to ~$Q(\bar{x})$ in ~$r$, the
query is rewritten into a new query ~${\cal T}(Q(\bar{x}))$; and
then the (ordinary) answers to ~${\cal T}(Q(\bar{x}))$ are retrieved
from $r$.

\begin{example} (example \ref{ex:repairs} continued) Consider the
query ~$Q:~ Supply(x,y,z)?$ about the items supplied together with their
associated information. In order to obtain the consistent answers,
the query ~${\cal T}(Q):~ Supply(x,y,z) \wedge
(Class(z,T_4) \rightarrow x  = C)$ is generated and posed to the original
database. The extra conjunct in it is the ``residue" obtained from the
interaction between the query and the constraint. Residues can be
obtained automatically \cite{ArenasBertossiChomicki:99}.
\boxtheorem
\end{example}

In general, ~${\cal T}$~ is an iterative operator. There are
sufficient conditions on queries and ICs for soundness,
completeness and termination of operator ${\cal T}$; and natural
and useful syntactical classes  satisfy those conditions. There
are some limitations though: ~ ${\cal T}$ can not be applied to
existential queries like ~$Q(X):~ \exists Y~ {\it
Supplies}(X,Y,It_1)?$. However, this query does have  consistent
answers at the semantic level. Furthermore, the methodology
presented in \cite{ArenasBertossiChomicki:99} assumes that the ICs
are (universal) constraints written in clausal form.

There are fundamental reasons for the limitations of the query
rewriting approach. If a FO query can be always rewritten into a
new FO query, then the problem of {\em consistent query answering}
(CQA) would have polynomial time data complexity. From the results
in this paper (see also \cite{janIPL}), we will see that CQA is
likely to have a higher computational complexity.

Notice that ${\cal T}$ is based on the interaction between the
queries and the ICs. It does not consider the interaction between
the ICs and the database instance. In this paper we concentrate
mostly on this second form of interaction. In particular, we wonder if we can
obtain an implicit and compact representation of the database
repairs.

Furthermore, the database seen  as a set of logical formulas plus
$IC$ is an inconsistent first order theory; and we know that such
an inconsistency can be detected and represented by means of an analytic
tableau.

An analytic tableau is a syntactically generated tree-like
structure that,
starting from a set of formulas placed at the root,
has all  its branches ``closed" when the initial  set of formulas is
inconsistent. This tableaux can show us how to repair
inconsistencies, because
 closed branches can be opened by removing  literals.

In this work, we show how to generate, close and open tableaux for database
instances with  their constraints; and we apply the notion of opening  to
characterize and represent by means of a tree structure all the repairs
of the original database. Finally, we sketch how this
representation could be used to retrieve consistent query answers.
At least at the theoretical level, the methodology introduced here
could be applied to any kind of first order queries and
constraints.

\section{Database Repairs and Analytic Tableaux}\label{se:DatabaseRepairsandAnalyticTableaux}

In order to use analytic tableaux to represent database repairs
and characterize consistent query answers, we  need a
special form of tableaux, suitable for representing database
instances and their integrity constraints.

Given a database instance $r$ and a finite set of integrity
constraints $IC$, we first compute the tableau,  $\TP(IC \cup r)$,
for $IC$ and $r$.  This tableau has as root node the set of
formulas~ $IC ~\cup~ r$. This tableau should be closed, that is
the  tableau has only closed branches,
 if and only if $r$ is inconsistent. By removing
database literals in every closed branch we can transform $r$ into
a consistent database instance and thus obtain a repair of the
database. For all this to work, we must take into account, when
computing the tableau, that $r$ represents a database instance and
not just a set of formulas, in particular, that the absence of
positive information means negative information, etc. (see section
\ref{sec:atdbi}). Next, we give a brief review of classical first
order analytic tableaux \cite{Beth:59,Smullyan:68,Fitting96}.

\subsection{Analytic tableaux}\label{se:prop}

The tableau of a set of
formulas is obtained by recursively {\em breaking down} the
formulas into subformulas, obtaining sets of sets of formulas.
These are the usual Smullyan's classes of formulas:

%\newpage \vspace*{-2cm}
$$\begin{array}{|c|cc||c|cc|}\hline
      \alpha            & \alpha_1 & \alpha _2 &      \beta        & \beta_1 & \beta_2 \\ \hline
f \wedge g              &   f      &   g       &   f \vee g        &    f    &    g    \\ \hline
\lnot(f \vee g)         & \lnot f  & \lnot g   & \lnot(f \wedge g) & \lnot f & \lnot g \\ \hline
\lnot(f \rightarrow g)  &   f      & \lnot g   & f \rightarrow g   & \lnot f &    g    \\ \hline
\end{array}$$
$$\begin{array}{|c|c||c|c|} \hline
     \gamma       & \gamma(p), p \mbox{ any constant} & \delta            & \delta(p),
     p \mbox{ a fresh constant} \\ \hline
(\forall x)f      & f [x/p]                            & (\exists x)f      & f [x/p]          \\ \hline
\lnot(\exists x)f & \lnot f [x/p]                      & \lnot(\forall x)f &\lnot f [x/p]     \\ \hline
\end{array}$$

\vspace{2mm} A tableaux prover produces a formula tree. An
$\alpha$-rule adds new formulas to  branches, a $\beta$-rule
splits the tableau and adds a new branch. Given a formula
$\varphi$, we denote by $\TP(\varphi)$ the tree produced by the
tableaux system. We can think of this tree as the set of its
branches, that we usually denote with $X,Y, \ldots$.

Notice
that the original set of constants in the language, in our case, $D$, is
extended with a set of new constants, $P$, the so-called Skolem functions or
parameters. These parameters, that we will denote by  $p, p_1, \ldots$, have
to be new at
the point of their introduction in the tree in the sense that they have not
appeared so far in the (same branch of the) tableau. When applying the
$\gamma$-rule, the parameter can be any of the old or new constants.

\ignore{ \comcs{It is the notion of saturated branch (very
classical in tableaux)!! You said:  ``We should also say that
whenever we write a branch $B = r \cup I$ it is understood that
$I$ is finished. '' Well I think this is not necessary: We only
use to write $B = r \cup I$ when $B$ is a tableau branch. But by
the definition, a tableau IS saturated!! (This can be shown)} }

 A
tableau branch is {\em closed} if it contains a formula and its
negation,~ otherwise it is {\em open}. Every open branch
corresponds to a model of the formula: If a branch $B \in
\TP(\varphi)$ is open and finished, then the set of ground atoms
on $B$ is a model of $\varphi$. If the set of initial formulas is
inconsistent, it does not have models, and then all branches (and
thus the tableau)  have to be closed. Actually, the  completeness
theorem for tableaux theorem proving \cite{Smullyan:68} states
that: ~ $F$ is a theorem iff $\TP(\{\neg F\})$ is closed.

The intuitive idea of finished branch, of one to which no tableaux
rule can be applied obtaining something new and relevant,  is
captured by means of the notion of \emph{saturated branch}: this
is a branch where all possible rules have been applied.

\begin{definition} \label{def:saturated} \em
A branch $B$ is \emph{saturated} iff it satisfies
\be
\item If ~$\neg\neg \varphi \in B$, then ~$\varphi \in B$
\item If ~$(\varphi \vee \psi) \in B$, then ~$\varphi \in B$ ~or~ $\psi \in B$
\item If ~$(\varphi \wedge \psi) \in B$, then ~$\varphi \in B$ ~and~ $\psi \in B$
\item If ~$\exists x \varphi \in B$, then ~$\varphi[c] \in B$ ~for some constant $c$
\item If ~$\forall x \varphi \in B$, then ~$\varphi[c] \in B$ ~for any constant
$c$.\footnote{If the language had function symbols, we would have
replace constants by ground terms in this definition.} \boxtheorem
\ee
\end{definition}

A branch is called \emph{Hintikka} if it is saturated  and not
closed \cite{Fitting96}. It is easy to see that a saturated branch
is Hintikka iff it does not contain any atomic formula $A$ and its
negation $\neg A$. From now on, tableaux branches will be assumed
to be saturated. Nevertheless, sometimes we talk about branches
even when they are partially developed only.

We consider $\TP$ not only as a theorem prover (or consistency
checker) for formulae but also as an application from (sets of)
formulas to trees which has some useful properties. Thus,
operations on tableaux can be defined on the basis of the logical
connectives occurring inside the formulas involved.

\begin{lem} \po \label{le:op}\em
 Let $\varphi$ and $\psi$ be any formulae. Then $\TP$ has the
 following properties.

 \be
 \item $\TP(\{\varphi \vee\psi\}) = \TP(\{\varphi\}) \cup \TP(\{\psi\})$
 \item $\TP(\{\varphi \wedge\psi\}) = \{X \cup Y: X \in \TP(\{\varphi\})$
 and Y $\in \TP(\{\psi\})\}$
  \item If $B\in \TP(\varphi\wedge \psi) $ then $B = B' \cup B''$ and
  $B'\in \TP(\varphi)$ and $B''\in \TP(\psi)$. \boxtheorem
 \ee
 \end{lem}

Property  3. follows directly from properties 1. and 2. The
properties in the lemma motivate  the following definition.

\begin{definition}\label{def:branches}\em Given tableaux $T$ and $T'$, each of them
identified with the set of its branches, the combined tableaux is~
 $T
\otimes T' = \{X \cup Y: X \in T \mbox{ and }Y \in T'\}.$
\boxtheorem
\end{definition}

\begin{remark} \label{rem:req}
The properties in lemma \ref{le:op} can be used to check whether a
formula $\varphi$ derives from a theory $A$. $A \models \varphi$
iff $(A \rightarrow \varphi)$ is a theorem, what will be proved if
we derive a contradiction from assuming $\neg(A \rightarrow
\varphi)$. Therefore we will have to compute $\TP(\{\neg(A
\rightarrow \varphi)\})$ and check for closure. Using the second
property, we will check $\TP(\{A\}) \otimes \TP(\{\neg\varphi\})$
for closure, allowing us to compute $\TP(A)$ only once for any
number of requests. \boxtheorem
\end{remark}

The following relationship between the open branches of the
tableaux for a formula and the its models has been shown, among
others by \cite{ABelletal:95,RischSchwind:94}.

\begin{theorem}\label{theo:modtab}\em
Let $B\in \TP(\{\phi\})$ be an open branch of the tableau for
$\phi$. Then there is a model $M$ of $\phi$, which satisfies $B$,
i.e. $B \subseteq M$. More precisely, there is Herbrand model of
$\varphi$ such that the ground atoms in $B$ belong to $M$.
\boxtheorem
\end{theorem}

\subsection{Representing database instances by tableaux} \label{sec:atdbi}

In database theory, we usually make the following
assumptions\footnote{Actually, it is possible to make all these
assumptions explicit and transform the database instance into a
first-order theory \cite{reiter:84}.}: ~(a) Unique Names
Assumption (UNA): If $a$ and $b$ are different constants in $D$,
then $a\neq b$ holds in $r$. ~(b) Closed World Assumption (CWA):
If $r$ is a database instance, then for any ground database atom
$P(c)$, if $P(c) \not\in r$, then $\neg P(c)$ holds for $r$, more
precisely, implicitly ~$\neg P(c)$ belongs to $r$.

In consequence, if we see the relational database as the set of
its explicit atoms plus its implicit negative atoms, we can always
repair the database by removing ground database literals.

When computing a tableau for a database instance $r$, we do not add explicitly
the formulas corresponding to the UNA and CWA, rather we keep them
implicit, but taking them into account when computing the tableau.
This means, for example,  that the presence on a tableau branch of a formula
$a=b$, for different constants $a, b$ in $D$, closes the branch.

Given a database $r$ and integrity constraints $IC$, we will
generate the tableau $\TP(\IC \cup r)$. Notice that every branch
$B$ of this tableau will be of the form $I \cup r$, where $I \in
\TP(\IC)$ (see lemma \ref{le:op}). $I$ is the ``$\IC$-part" of the
branch.

Notice also that a tableau for $\IC$ only will never be closed,
because $\IC$ is consistent. The same happens with any tableau for
$r$. Only the combination of $r$ and $\IC$ may produce a closed
tableau.

$\TP(\IC \cup r)$ is defined as in section \ref{se:prop}, but we
still have to define the closure conditions for tableaux
associated to database instances. Before, we present some
motivating examples.

\begin{example}\label{ex:SupplyClass} (example \ref{ex:repairs}
continued) ~In this case, $\TP(\IC \cup r)$ is the  tree in figure
\ref{fig1}.
\begin{figure}[hbt]
\hspace*{-1.5cm}
\ptbegtree
    \ptbeg
 \npile{\forall x, y, z (Supply(x,y,z) ~\wedge~ Class(z,T_4)~\rightarrow~ x=C)\\
Supply(D, D_2, It_2)\\
        Supply(C, D_1, It_1)\\
        Class(It_1, T_4)\\
        Class(It_2, T_4)}
        \ptbeg \ptnode{$Supply(C, D_1, It_1) ~\wedge~ Class(It_1, T_4)
~\rightarrow~ C=C$}
 \ptbeg \ptnode{$Supply(D, D_2, It_2) ~\wedge~ Class(It_2, T_4) ~\rightarrow~ D=C$}
          \npile{\neg Supply(C, D_1, It_1)\\
          \dots \\
                \times}
\npile{~\neg Class(It_1, T_4)\\
        \dots \\
                \times}
        \ptbeg \ptnode{$C=C$}
                \npile{~\neg Supply(D, D_2, It_2)\\
                \times}
               \npile{~~\neg Class(It_2, T_4)\\
               \times}
              \npile{~D=C\\
               \times} \ptend
\ptend \ptend \ptend
\ptendtree\\
\begin{center}%\usebox{\TeXTree}
\caption{Tableau for Example \ref{ex:SupplyClass}} \label{fig1}
\end{center}
\end{figure}
The last branch is closed because $D = C$ is false in the database
(alternatively, because $D \neq C$ is implicitly in the database).
We can see that $\TP(\IC \cup r)$ is closed. $r$ is inconsistent
wrt $IC$. The nodes  ~$(\nit{Supply}(C, D_1, \nit{It}_1) ~\wedge~
\nit{Class}(It_1, T_4)$ $\rightarrow~ C=C)$~ and $(\nit{Supply}(D,
D_2, It_2) ~\wedge~ \nit{Class}(It_2, T_4) ~\rightarrow~ D=C)$~
are obtained by applying the $\gamma$-rule to $\forall x, y, z
(\nit{Supply}(x,y,z) ~\wedge~ \nit{Class}(z,T_4)~\rightarrow~
x=C)$. Application of the $\beta$-rule to $(\nit{Supply}(D, D_2,
It_2) ~\wedge~ \nit{Class}(It_2, T_4) ~\rightarrow~ D=C)$ produces
the same subtree for all three leaves: $\neg \nit{Supply}(C, D_1,
It_1)$,  $\neg \nit{Class}(It_1, T_4)$ and $C=C$. In the figure,
we indicate this subtree by ``\dots''. We will see later (see
section \ref{subse:OpeningTableaux}) that, in some cases, we can
omit the development of subtrees that should develop under
branches that are  already closed.  Here we can omit the explicit
further development of the subtree from the first two leftmost
branches, because these branches are already closed. \boxtheorem
\end{example}

In tableaux with equality, we  need extra rules. We will assume
that we can always introduce equalities of the form $t = t$, for a
term $t$, and that we can replace a term $t$ in  a predicate $P$
by $t'$ whenever $t=t'$ belongs to the same tableau branch
(paramodulation, \cite{Fitting96}).  It will be simpler to define
the closure rules for database tableaux, if we skolemize
existential formulas before developing the tableau
\cite{FittingJAR:88}. We assume from now on that all integrity
constraints are skolemized by means of a set of Skolem constants
(the parameters in $P$) and new function symbols.

\begin{example} \label{ex:refIC2}
Consider the referential $IC:~ \forall x~(P(x) \rightarrow \exists
y~Q(x,y))$, and the inconsistent database instance ~$r = \{P(a),
Q(b,d)\}$, for $a, b, c \in D$. With an initial skolemization, we
can develop the following tableau $\TP(\IC \cup r)$. In this
tableau, the second branch closes because $Q(a,f(a))$ does not
belong to the database instance. There is no $x$ in the active
database domain, such that $r$ contains $Q(a,x)$. Implicitly, by
the CWA, $r$ contains then $\neg Q(a,x)$ for any $x$. Hence the
branch containing $Q(a,f(a))$ closes and $r$ is inconsistent for
$IC$.

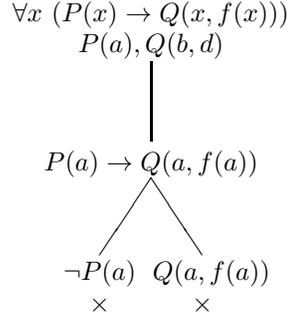
\begin{center}
\ptbegtree
    \ptbeg
 \npile{\forall x~(P(x) \rightarrow Q(x,f(x)))\\
P(a), Q(b,d)} \ptbeg
          \ptnode{$P(a) \rightarrow Q(a,f(a))$}
          \npile{\neg P(a)\\
          \times}
 \npile{~~Q(a,f(a))\\ \times}
               \ptend \ptend
\ptendtree
\end{center}
\end{example}

\begin{example}\label{ex:cons}
Consider the inconsistent database $r_1 = \{Q(a), Q(b)\}$ wrt the $IC$:
$\exists x~P(x)$. After having skolemized $\exists x ~ P(x)$ into $P(p)$, a
 tableau proof for the inconsistency is the following

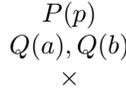
\begin{center}
 $P(p)$

$Q(a), Q(b)$

$\times$
\end{center}

This branch closes because there is no $x$ in $D$ such that $P(x)
\in r$ and therefore $\neg P(x)$ belongs to $r$ for any $x$ in
$D$.  $P(p)$ cannot belong to this database. \boxtheorem
\end{example}

\begin{example}\label{ex:newexa}
Let us now change the database instance in example \ref{ex:cons}
to $r_2 = \{P(a), P(b)\}$, keeping the integrity constraint. Now,
the database is consistent, and we have  the following tableau
$\TP(\IC\cup r_2)$:

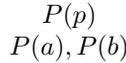
\begin{center}
 $P(p)$

$P(a), P(b)$
\end{center}

This time we do not want the tableau to close, and thus sanctioning the
inconsistency of the database. The reason is that we could make $p$
take any of the values in the active domain $\{a, b\} \subseteq D$
of the database. \boxtheorem
\end{example}

A similar situation can be found in a modified version of example \ref{ex:refIC2}.

\begin{example}\label{exa:nnewexa} Change
the database instance in example \ref{ex:refIC2} to $\{P(a),
Q(a,d)\}$. Now it is consistent wrt the same IC. We obtain

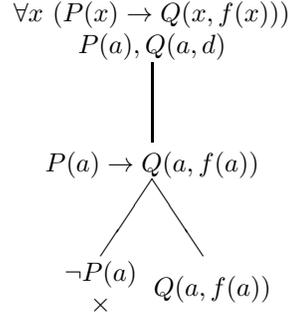
\begin{center}
\ptbegtree
    \ptbeg
 \npile{\forall x~(P(x) \rightarrow Q(x,f(x)))\\
P(a), Q(a,d)} \ptbeg
          \ptnode{$P(a) \rightarrow Q(a,f(a))$}
          \npile{\neg P(a)\\
          \times}
 \npile{~~Q(a,f(a))}
              \ptend \ptend
\ptendtree
\end{center}

Now we do not close the rightmost branch because we may define  $f$
as a function from the active domain into itself that makes
$Q(a,f(a))$ become a member of the database, actually by defining
$f(a) = d$. \boxtheorem
\end{example}

\begin{example} \label{ex:neg}
Consider $IC:~ \exists x~ \neg P(x)$ and the consistent database
instance $r = \{P(a)\}$. The tableau $\TP(\IC \cup r)$ after
skolemization of $IC$ is:
\begin{center}
$\neg P(p)$

$P(a)$
\end{center}

This tableau cannot be closed, because $p$ must be a new
parameter, not occurring in the same branch of the tableau and it
is not the case that $P(p) \in r$ (alternatively, we may think of
$p$ as a constant that can be defined as any element in $D
\setminus \{a\}$, that is in the complement of the active domain
of the database). \boxtheorem
\end{example}

In general, a tableau branch closes whenever it contains a formula
and its negation. However, in our case, it is necessary to take
into account that not all literals are explicit on branches due to
the UNA and CWA. The following definition of closed branch
modifies the standard definition, and considers those assumptions.

\begin{definition}\label{de:dataclos}\em
Let $B$ be a  tableau branch for a database instance $r$ with
integrity constraints $IC$, say $B= I \cup r$. $B$ is closed iff one
of the following conditions holds: \be
\item
$a=b \in B$ for different constants $a, b$ in $D$.

\item \begin{enumerate} \item $P(\bar{c})\in I$ and $P(\bar{c})\not\in r$,
for a ground tuple $\bar{c}$ containing elements of $D$ only.

\item
$P(\bar{c})\in I$ and there is no substitution $\sigma$ for the
parameters in $\bar{c}$ such that $P(\bar{c})\sigma \in
r$.\footnote{A substitution is given as a pair $\sigma =(p,t)$,
where $p$ is a variable (parameter) and $t$ is a term. The result
of applying $\sigma$ to formula $F$, noted $F\sigma$, is the
formula obtained by replacing every occurrence of $p$ in $F$ by
$t$.}
\end{enumerate}
\item  $\neg P(\bar{c})\in I$ and $P(\bar{c})\in r$ for a ground tuple
$\bar{c}$ containing elements of $D$ only.

\item $\varphi \in B$ and $\neg\varphi \in B$, for an arbitrary formula
$\varphi$.

\item $\neg t = t ~\in B$ for any term $t$. \boxtheorem
\ee
\end{definition}

Condition 1. takes UNA into account. Notice that it is restricted
to database constants, so that it does not apply to  new
parameters\footnote{That is, elements of $P$ are treated as null
values in Reiter's logical reconstruction of relational databases
\cite{reiter:84}.}. Condition 2(a) takes CWA into account.
Alternative condition 2(b) (actually it subsumes 2(a)) gives an
account of examples \ref{ex:refIC2},   \ref{ex:cons},
\ref{ex:newexa}, and \ref{exa:nnewexa}.

In condition 3. one might miss a second alternative as in
condition 2., something like ``$\neg P(\bar{c})\in I$ for a ground
tuple containing Skolem symbols, when there is no way to define
them considering elements of $D \setminus \nit{Act}(r)$ in such a
way that $P(\bar{c})\not\in r$". This condition can be never
satisfied because we have an infinite database domain $D$, but a
finite active domain $\nit{Act}(r)$. So, it will never apply. This
gives an account of example \ref{ex:neg}. Conditions 4. and 5. are
the usual closure conditions. Conditions  2(a) and 3. are special
cases of 4.

Now we can state the main properties of tableaux for database instances and
their integrity constraints.

\begin{proposition}\em
For a database instance $r$ and integrity constraints $IC$, it
holds:
\begin{enumerate}
\item $r$ is inconsistent wrt to $IC$ ~iff~   the tableau
$\TP(\IC \cup r)$ is closed (i.e. each of its branches is closed).
\item
 $\TP(\IC  \cup r)$ is  closed iff ~$r$ does not satisfy $\IC$ (i.e. $r \not \models \IC$). \boxtheorem
 \end{enumerate}
\end{proposition}

\subsection{Opening tableaux}\label{subse:OpeningTableaux}

The inconsistency of a database ~$r$~  wrt $\nit{IC}$~ is
characterized by a tableau ~$\TP(\IC \cup  r)$ which has only
closed branches. In order  to obtain a repair of $r$, we may
remove  the literals  in the branches which are ``responsible" for
the inconsistencies, even implicit literals  corresponding to the
CWA. Every branch which can be ``opened" in this way will possibly
yield a repair. We can only repair inconsistencies due to literals
in $r$. We cannot  remove literals in $I$ because, according to
our approach, integrity constraints are rigid, we are not willing
to give them up; we only allow changes in the database instances.
We cannot suppress equalities $a=b$ neither built-in predicates.

\begin{remark} \label{rem:ops}
According  to  Definition \ref{de:dataclos}, we can  repair
inconsistencies due only to cases 2. and 3. More precisely, given
a closed branch $B$ in $\TP(\IC \cup r)$:
\begin{enumerate}
\item If $B$ is closed because of the CWA,
 it can be opened by inserting $\sigma P(\bar{c})$ into $r$, or, equivalently, by removing the implicit
 literal $\neg \sigma P(\bar{c})$ from $r$ for any substitution $\sigma$ from the parameters into $D$
 (case 2(b) in Def. \ref{de:dataclos}).
 \item If $B$ is closed because of
contradictory literals $\neg P(\bar{c})\in I$ and $P(\bar{c}) \in
r$, then it can  be opened by removing $P(\bar{c})$ from $r$ (case
3 in Def. \ref{de:dataclos}) . \boxtheorem
\end{enumerate}
\end{remark}

\begin{example} \label{exa:Supply3} (example \ref{ex:SupplyClass} continued)
The tableau has 9 closed branches: (we display  the
literals within the branches only)

\( \begin{array}{ccc}
  B_1 & B_2 & B_3 \\
   Supply(C, D_1,It_1) & ~~~~Supply(C, D_1,It_1) & ~~~~Supply(C, D_1,It_1) \\
Supply(D, D_2, It_2) & Supply(D, D_2, It_2) & Supply(D, D_2, It_2)
\\ Class(It_1, T_4) & Class(It_1, T_4) & Class(It_1, T_4) \\
Class(It_2, T_4) & Class(It_2, T_4) & Class(It_2, T_4) \\ \neg
Supply(C, D_1,It_1) & \neg Supply(C, D_1,It_1) & \neg Supply(C,
D_1,It_1) \\ \neg Supply(D, D_2, It_2) & \neg Class(It_2, T_4) & D
= C\\
   &   &  \\
\end{array}
\)

\( \begin{array}{ccc}
  B_4 & B_5 & B_6 \\
   Supply(C, D_1,It_1) & ~~~~Supply(C, D_1,It_1) & ~~~~Supply(C, D_1,It_1) \\
Supply(D, D_2, It_2) & Supply(D, D_2, It_2) & Supply(D, D_2, It_2)
\\ Class(It_1, T_4) & Class(It_1, T_4) & Class(It_1, T_4) \\
Class(It_2, T_4) & Class(It_2, T_4) & Class(It_2, T_4) \\ \neg
Class(It_1, T_4) & \neg Class(It_1, T_4) & \neg Class(It_1, T_4)
\\ \neg Supply(D, D_2, It_2) & \neg Class(It_2, T_4) & D = C\\
   &   &  \\
\end{array}
\)

\( \begin{array}{ccc}
  B_7 & B_8 & B_9 \\
   Supply(C, D_1,It_1) & ~~~~Supply(C, D_1,It_1) & ~~~~Supply(C, D_1,It_1) \\
Supply(D, D_2, It_2) & Supply(D, D_2, It_2) & Supply(D, D_2, It_2)
\\ Class(It_1, T_4) & Class(It_1, T_4) & Class(It_1, T_4) \\
Class(It_2, T_4) & Class(It_2, T_4) & Class(It_2, T_4) \\ C = C &
C=C & C= C  \\ \neg Supply(D, D_2, It_2) & \neg Class(It_2, T_4) &
D = C\\
   &   &  \\
 \end{array}
\)

The first four tuples in every branch correspond to the initial
instance $r$. Each branch $B_i$ consists of an $I$-part and the
$r$-part, say $B_i = r \cup I_i$. And we have

 \( \begin{array}{ccc}
  I_1 & I_2 & I_3 \\
   \neg Supply(C, D_1,It_1) & ~~~~\neg Supply(C, D_1,It_1) & ~~~~\neg Supply(C, D_1,It_1) \\
\neg Supply(D, D_2, It_2) & \neg Class(It_2, T_4) & D = C\\
   &   &  \\
\end{array}
\)

\( \begin{array}{ccc}
  I_4 & ~~~~I_5 & ~~~~~~I_6 \\
   \neg Class(It_1, T_4) & ~~~~\neg Class(It_1, T_4) & ~~~~~~~~\neg Class(It_1, T_4) \\
\neg Supply(D, D_2, It_2) & ~~~~\neg Class(It_2, T_4) & ~~~~~~~~D
= C\\
   &   &  \\
\end{array}
  \)

 \( \begin{array}{ccc}
  I_7 & ~~~~~~~I_8 & ~~~~~~~I_9 \\
   C = C & ~~~~~~~~C=C & ~~~~~~~~C= C  \\
\neg Supply(D, D_2, It_2) & ~~~~~~~\neg Class(It_2, T_4) &
~~~~~~~~D = C\\
   &   &  \\
\end{array}
\)

In order to open  this closed tableau, we can remove literals in
the closed branches. Since a tableau is open whenever it has an
open branch, each opened branch of the closed tableau might
produce one possible transformed open tableau. Since we  want to
modify the database $r$, which should become consistent, we
should  try to remove a minimal set of literals in the $r$-part of
the  branches in order to open the tableau. This automatically
excludes branches $B_3, B_6$ and $B_9$, because they close due to the literals $D = C$,
which do not correspond to database literals, but come from the constraints.

In this example we  observe that the sets of database literals of
some of the $I_j$ are included in others. Let us denote by $I'_j$
the set of  literals in $I_j$  that are database literals (i.e.
not built-in literals), e.g. $I'_1 = I_1, I'_7 =  \{\neg
\nit{Supply}(D,D_2,$ $It_2)\}$. We have then $I'_1 \supset I'_7$,
$I'_2 \supset I'_8$, $I'_3 \supset I'_9$, $I'_4 \supset I'_7$,
$I'_5 \supset I'_8$, $I'_6 \supset I'_9$. This shows, for example,
that in order to open $B_1$,  we have to remove from $r$ a
superset of the set of literals  that have to be removed from $r$
for opening $B_7$. Hence, we can decide that the branches whose
database part contains the database part of another branch can be
ignored because they will not produce any (minimal) repairs. This
allows us not to consider $B_1$ through $B_6$ in our example, and
$B_7$ and $B_8$ are the only branches that can lead us to repairs.
\boxtheorem
\end{example}

The following lemma tells us that we can ignore branches with
subsumed $I$-parts, because those branches cannot become repairs.

\begin{lem} \po \label{le:openings}\em
 If ~$r'' \subseteq r' \subseteq r$, then  ~$r' \leq_{r}   r''$.
 \boxtheorem
\end{lem}

\ignore{
 \comlb{I guess you mean $r' \leq_r r''$ on the RHS,
because otherwise, it is easy to get a contradiction taking $r' =
r''$. Again, if we need something like this, would it not be
better to push it to the appendix before the place where it is
needed? Independently of this, I have the impression that we need
something similar, but different (if we want to capture the
intuition expressed before the lemma), actually related to closed
branches,  e.g. the following: After having defined openings (next
definition), we could state and prove that, for $B_1 = I_1 \cup r,
~B_2 = I_2 \cup r, ~I'_1 \subseteq I'_2,~ r_1 = op(B_1), r_2 =
op(B_2) \Rightarrow r_1 \leq_r r_2$. }

\comcs{It was wrong and I corrected it. } }

 Moreover, as
illustrated above, where the tableau tree is shown, sometimes we
can detect possible subsuming branches without fully developing
the tableau. In example \ref{ex:SupplyClass} the first formula has
been split by a tableau rule and we have already closed two
branches. When we apply another rule, we know then, that the
branch $C=C$, which is not closed yet,  will be not be closed or
will be closed by a subset of the database literals appearing in
the first two branches.

\begin{definition}\label{de:opening}\em
 Let $B= I \cup r$ be a closed branch of the tableau  $\TP(\IC
\cup r)$.
\begin{itemize}
\item[(a)] If $I$ is  not closed, i.e the branch is closed due to
database literals only, we say that $B$ is {\em data closed}.
\item[(b)] Let
 $B= I \cup r$ be a data closed
branch in the tableau  $\TP(\IC \cup r)$, we define $\opb := (r
\setminus L(B)) \cup K(B)$, where
\begin{enumerate} \item $L(B) = \{l ~|~ l \in r \mbox{ and } \neg
l \in I\}$ \item  $K(B) = \tau \{l ~|~ l \mbox{ is a ground atom
in} I \mbox{ and there is no substitution } \sigma \mbox{ such}$
$\mbox{that }$ $ l\sigma \in r\}$, where $\tau$ is any
substitution of the parameters into $D$.
\end{enumerate}
\item[(c)] An instance $r'$ is called an {\em opening} of ~$r$ iff
$r' = \opb$ for a data closed branch $B$ in $\TP(\IC \cup r)$.
\boxtheorem
\end{itemize}
\end{definition}

If the branch $B$ is clear from the context, we simply write $r' =
(r \setminus L) \cup K$. If no parameters have been introduced in
the branch, then we do not need to consider the substitutions
above. In this case, for an opening $I \cup r'$ of a branch $I \cup r$
it holds:(a) If  $P(\bar{c}) \in I$ and  $P(\bar{c})\not \in  r$,
then   $P(\bar{c}) \in r'$. (b) If  $\neg P(\bar{c}) \in I$ and
$P(\bar{c}) \in r$,  then  $P(\bar{c})\not \in r'$. Notice that we
only open branches which are closed because of conflicting
database literals.

When $r \models \IC$, then $\TP(\IC \cup r)$ will have (finished) open
branches $B$. For any of those branches $\opb$ can be defined exactly
as in Definition \ref{de:opening}. It is easy to verify that in this case
$\opb$ coincides with the original instance $r$.

\begin{proposition} \label{prop:model}\em
Let ~$r'$ be an opening of $r$. Then ~$r'$ is consistent with
$\IC$, i.e. ~$r' \models \IC$. \boxtheorem
\end{proposition}

\begin{example}\label{ex:models}
Consider  $r= \{P(a), Q(a), R(b)\}$ and $\IC =
\{\forall x(P(x) \rightarrow Q(x)\}$. Here $r \models \IC$ and $\TP(\IC \cup
r)$ is
\begin{center}
\ptbegtree
    \ptbeg
     \npile{P(a), Q(a), R(b)\\P(x) \rightarrow Q(x)}  \ptbeg \npile{\neg P(b)~~~\\}
      \npile{\neg P(a)~~~\\ \times~~~\\B_1}
      \npile{~~~Q(a)\\ B_2} \ptend \npile{~~~~Q(b)\\ \times\\B_3}
      \ptend \ptendtree
      \end{center}
      The first branch, $B_1$, is closed and $\nit{op}(B_1) =
      \{Q(a), R(b)\}$ that satisfies $\IC$. The second branch, $B_2$, is
      open and $\nit{op}(B_2) = r$. The third branch, $B_3$, is closed
      and $\nit{op}(B_3) = \{P(a), Q(a), Q(b), R(b)\}$ that satisfies
      $\IC$. Notice that we could further develop the last node there,
      obtaining the same tree that is hanging from $\neg P(b)$
      in the tree on the LHS. If we do this, we obtain closed branches
      $B_4, B_5$, with $\nit{op}(B_4) = \{Q(a), Q(b), R(b)\}$, and
      $\nit{op}(B_5) = \{P(a), Q(a), Q(b), R(b)\}$. With these last two
      openings we do not get any closer to $r$ than with $\nit{op}(B_3)$,
      that is still not as close to $r$ as the only repair, $r$, obtained
      with branch $B_2$. \boxtheorem
\end{example}

\begin{example}\label{ex:models2}
Consider  $\IC$ as in example \ref{ex:models}, but now $r= \{P(a), R(b)\}$,
that does not satisfy $\IC$. $\TP(\IC \cup r)$ is
\begin{center}
\ptbegtree
    \ptbeg
     \npile{P(a), R(b)\\P(x) \rightarrow Q(x)}  \ptbeg \npile{\neg P(b)~~~\\}
      \npile{\neg P(a)~~~\\ \times~~~\\B_1}
      \npile{~~~Q(a)\\ ~~~\times\\B_2} \ptend \npile{~~~~Q(b)\\ \times\\B_3}
      \ptend \ptendtree
      \end{center}
      For the first branch $B_1$, we obtain $\nit{op}(B_1) =
     \{R(b)\}$, that is a repair. Branch $B_2$ gives $\nit{op}(B_2)
     = \{P(a),  R(b), Q(a)\}$, the other repair.

     For the closed branch $B_3$ we have $\nit{op}(B_3) =
     \{P(a), Q(b), R(b)\}$.
     This is not a model of $\IC$, apparently contradicting Proposition \ref{prop:model},
     in particular, it is not a repair of $r$.
     If we keep
      developing node $Q(b)$ exactly as $\neg P(b)$ on the LHS, we obtain
      extended (closed) branches, with associated instances
      $\{Q(b), R(b)\}$ and $\{P(a), Q(a), Q(b),
      R(b)\}$. Both of them satisfy $\IC$, but are non minimal; and then they
      are not repairs of $r$. This example shows the importance of having
      the open and closed branches (maybe not explicitly) saturated (see Definition
      \ref{def:saturated}).
      \boxtheorem
      \end{example}

We can see that every opening is related to a possibly non
minimal repair of the original database instance\footnote{Strictly speaking, we should not say
``non minimal repair", because repairs are minimal by
definition. Instead, we should talk of database instances that differ from the
original one and satisfy the ICs. In any case, we think there should be no
confusion if we relax the language in this sense.}. For repairs, we
are only interested in ``minimally'' opened branches, i.e. in open
branches which are as close as possible to $r$.  In consequence, we may
define a {\em minimal opening} $r'$ as an opening such that $r \Delta r'$ is
minimal under set inclusion.

Openings of $r$
are obtained by deletion of literals from $r$, or, equivalently,
by deletion/insertion of atoms from/into $r$. In order to obtain
minimal repairs, we have to make a minimal set of changes,
therefore we do not keep openings associated to an $r''$, such
that $r'\Delta r ~\subsetneqq~ r''\Delta r$, where $r'$ is
associated to another opening. We will show subsequently that
these are the openings where $L$ and $K$ are minimal in the sense
of set inclusion wrt all other openings in the same tree.

The following theorem establishes a relationship between the order
of repairs defined in Definition \ref{de:close} and the set
inclusion of the database atoms that  have been inserted or
deleted when opening a database instance.

\begin{lem}\po \label{le:LK} \em  For any opening $r' = (r\setminus L)\cup K$,
 we have $r\Delta r' = L \cup K$. \boxtheorem
\end{lem}

\begin{proposition}\label{theo:leqr}\em
Let $r_1 = (r\setminus L_1)\cup K_1$ and $r_2 = (r\setminus
L_2)\cup K_2$. Then   $r_1$ is closer to $r$ than $r_2$, i.e. $r_1
\leq_{r} r_2$ iff  $L_1 \subseteq L_2$ and $K_1 \subseteq K_2$.
\boxtheorem
\end{proposition}

\begin{theorem}\label{theo:repairop}\em
Let ~$r$ be an inconsistent database wrt $IC$. Then  $r'$ is a
repair of $r$  iff there is an open branch $I$ of $\TP(\IC)$, such
that $I\cup r$ is closed and  $I \cup r'$ is a minimal opening of
$I \cup r$ in $\TP(\IC \cup r)$. \boxtheorem
\end{theorem}

\begin{example}
(example \ref{exa:Supply3} continued) $\TP(\IC\cup r) $ has two
minimal openings:
\[\begin{array}{cc}
r'_7 & ~~~~r'_8\\ Supply(C,D_1, I_1) & ~~~~Supply(C, D_1, I_1)\\
Class(I_1, T_4) & ~~~~Class(I_1, T_4)\\ Class(I_2, T_4) &
~~~~Supply(D, D_2, I_2)\\
\end{array}\]
The rightmost closed branch cannot be opened because it is closed by
the atom $D=C$ which is not a database predicate. \boxtheorem
\end{example}

\section{Repairs, Knowledge Base Updates and Complexity}\label{se:Complexity}

Our definition of  repairs is based on a minimal distance function
as used by Winslett for knowledge base update \cite{Winslett:88}.
More precisely, Winslett in her ``possible models approach''
defines the knowledge base change operator $\circ$ for the update
of a propositional knowledge base $K$ by a propositional formula
$p$ by
$$Mod(K\circ p) = \hspace{-2mm} \bigcup_{m\in Mod(K)} \hspace{-3mm}\{m'\in Mod(p): m\triangle
m' \in Min_{\subseteq} (\{m\triangle m': m' \in Mod(p)\})\}$$

In \cite{EiterGottlob:92}, Eiter and Gottlob present
complexity results for propositional knowledge base revision and
update.
According to these results, Winslett's update operator is on the
second level of the polynomial  hierarchy in the general case
(i.e. without any syntactic restriction on the propositional
formulas): the problem of deciding whether a formula $q$ is a
logical consequence of  the update by $p$ of a knowledge base $T$
is $\Pi^P_2-$complete. \\

\begin{tabular}{|c|c|c|c|c|}
   \hline
  Update & General case & General case & Horn & Horn \\  \hline
     & arbitrary p & $\parallel p \parallel \leq k$ & arbitrary p & $\parallel p \parallel \leq k$
     \\ \hline
  $T\circ p \rightarrow q $ & $\Pi^P_2-$complete & co-NP-complete & co-NP-complete &
  $O(\parallel T\parallel \cdot \parallel q\parallel)$  \\ \hline
\end{tabular}
\\

In the above table, we resume the results
reported in \cite{EiterGottlob:92}. The table contains five
columns. In the general case (columns two and three), $T$ is a
general propositional knowledge base. In the Horn-case (columns
four and five), it is assumed that $p$ and $q$ and all formulas in
$T$ are conjunctions of Horn-clauses. Columns two and four account
for cases where no bound is imposed on the length of the update
formula $p$, while columns three and five describe the case where
the length of $p$ is bounded by a constant $k$. The table
illustrates that the general problem in the worst case (arbitrary
propositional formulas without bound on the size) is intractable,
whereas it becomes very well tractable (linear in the size of $T$
and query $q$) in the case of Horn formulas with bounded size.

How are these results related to CQA? If $r$ is a database
which is  inconsistent with respect to the set of integrity
constraints $\IC$, the derivation of a consistent  answer to a
query $Q$ from $r$ corresponds to the derivation of $Q$ from the
data base $r$ updated by the integrity constraints  $IC$. Hence,
the (inconsistent) knowledge base instance $r$, which is just a
conjunction of literals, corresponds to the propositional
knowledge base $T$. The
 integrity constraints $IC$ correspond to  the update formula $p$
 And deriving an answer to query $Q$ from $r$ (and $IC$) corresponds
 to the derivation of $Q$ from $r$ updated by $IC$.

Update is defined for propositional formulas.  Update is defined
by means of models of the knowledge base $r$ and the update
formula $IC$.  In our case, $r$
   is a finite conjunction of
grounded literals, i.e. $r$ is a propositional Horn formula. The
update formulas however (integrity constraints $IC$) are FO formulas.
However, the Herbrand universe of the
database is a finite set of constants. Therefore, we can consider
instead of  $\IC$ the finite set of instantiations of the formulas
in $\IC$ by database constants. Let us denote the conjunction of
these instantiations by $\nit{ic}$. Note that $\nit{ic}$ is Horn whenever all
formulas in $\IC$ are Horn, what is common for database ICs.

It is then easy to see that the following relationship holds
between update and repairs and CQA. It follows straightforwardly
from the definitions of repairs and update.

\begin{theorem}\em
Given a database instance $r$ and a set of integrity constraints
$\IC$ with their propositional database representation $\nit{ic}$:
\\
(a) $r'$ is a repair of ~$r$ wrt $\IC$ iff  ~$r' \in Mod(r\circ
\nit{ic})$.\\ (b) If $Q$ is a query, $\bar{t}$ is a consistent
answer to $Q$ wrt $\IC$ iff every model of ~$r\circ \nit{ic}$~ is
a model of $Q(\bar{t})$, i.e. $Mod(r\circ \nit{ic}) \subseteq
Mod(Q(\bar{t}))$. \boxtheorem
\end{theorem}

In consequence, the results given by Eiter and Gottlob apply
directly to CQA.

The number of branches of a fully developed tableaux is very high:
in the worst case, it contains $o(2^n)$ branches where $n$ is the
length of the formula. Moreover, we have to find minimal elements
within this exponential set, what increases the complexity.
Theorem \ref{theo:leqr} tells us that we do not need  to compare
the entire branches but only parts of them, namely the literals
which have been removed
 in order to open the tableau. This reduces the size of the
sets we have to compare, but not their number. Let us reconsider
in example \ref{ex:SupplyClass} the point just before applying the
tableaux rule which develops formula $Supply(D, D_2, I_2) ~\wedge~ Class(I_2,
T_4) ~\rightarrow~ D=C$. As we pointed out in the discussion of
example \ref{ex:SupplyClass}, under some conditions, it is
possible to avoid the development of closed branches because we
know in advance, without developing them, that they will not be minimal.

\begin{example}\label{ex:SupplyClass2} (example \ref{ex:SupplyClass}
continued) ~In this case, $\TP(\IC \cup r)$ is the  tree in Figure
\ref{fig2}. This tree has two closed branches, $B_1$ and $B_2$,
and one open branch $B_3$. Each of these branches will receive an
identical subtree due to the application of the tableaux rules to
the formulas not yet developed on the tree, namely
$(Supply(D,D_2,It_2) ~\wedge~ Class(It_2,T_4)~\rightarrow~ D=C)$.
We know at this stage of the development that $B_1$ is closed due
to $\neg Supply(C, D_1, It_1)$ and $B_2$ is closed due to $\neg
Class(It_1, T_4)$; $B_3$ is not closed. \boxtheorem
\end{example}

In this example, we can see that if we
further develop the tree, every $B_i$
will have the same sets of sub-branches, say $L_1$,  $L_2$, \dots,
where $L_i$ is a set of literals.
The final fully developed
tableau will then consist of the branches $B_1 \cup L_1$, $B_1
\cup L_2$, \dots, $B_2 \cup L_1$, $B_2 \cup L_2$, \dots  $B_3 \cup
L_1$, $B_3 \cup L_2$, \dots, \dots.
If the final tableau is
closed, since $B_3$ is not closed, every $B_3 \cup L_j$ will be
closed due to literals within $L_j$, say $K_j$.

We have then two cases: either the literals in $K_j$ close due to
literals in $r$ (which is the original inconsistent database
instance) or they close due to literals in the part of $B_3$ not
in $r$. In the first case, these literals from $K_j$ will close
every branch of the tree (also $B_1$ and $B_2$). Since $B_1$ and
$B_2$ were already closed, they will be closed due to a set of
literals that is strictly bigger than before, and therefore they
will not produce minimally closed branches (and no repairs). In
this situation, those branches can immediately be ignored and not
further developed. This can considerably reduce the size of the
tableau. In this example, at the end of the development, only
$B_3$ will produce repairs (see example \ref{ex:SupplyClass}).

In the second case, the literals in $K_j$ close due to literals in
the part of $B_3$ that are not in $r$. If these literals are not
database literals (we have called them built-in predicates), the
branch cannot be opened, we cannot repair inconsistencies that are
not due to database instances. Then, we only have to consider the
case of database literals that are not in $r$.

\begin{figure}[hbt]
\hspace*{-1.5cm}
\begin{center}
 \ptbegtree
    \ptbeg
 \npile{\forall x, y, z (Supply(x,y,z) ~\wedge~ Class(z,T_4)~\rightarrow~ x=C)\\
Supply(D, D_2, It_2)\\
        Supply(C, D_1, It_1)\\
        Class(It_1, T_4)\\
        Class(It_2, T_4)}
        \ptbeg \ptnode{$Supply(C, D_1, It_1) ~\wedge~ Class(It_1, T_4)
~\rightarrow~ C=C$}
 \ptbeg \ptnode{$Supply(D, D_2, It_2) ~\wedge~ Class(It_2, T_4) ~\rightarrow~
 D=C$}
          \npile{\neg Supply(C, D_1, It_1)\\
         \times\\
         B_1}
\npile{~~\neg Class(It_1, T_4)\\
         \times\\
         B_2}
        \npile{~~$C=C$\\
        B_3}
  \ptend \ptend \ptend
\ptendtree\\
\end{center}
\begin{center}
\caption{} \label{fig2}
\end{center}
\end{figure}

Since $B_3$ is open, those literals are negative literals (in the
other case, $B_3$ would not have been open, due to condition 2. in
Definition \ref{de:dataclos}). This is the only situation where
the sub-branches which are closed at a previous point of
development may still become minimal. In consequence, a reasonable
heuristics will be to suspend the explicit development of already
closed branches unless we are sure that this case will not occur.

\section{Consistent Query Answering}\label{se:QueryAnswering}

In order to determine consistent answers to queries, we can also
use, at least at the theoretical level,  a tableaux theorem prover
 to produce $\TP(\IC \cup r)$ and its openings. Let us
denote by $\nit{op}(\TP(\IC \cup r))$ the tableau $\TP(\IC \cup
r)$, with its minimal openings: All branches which cannot be
opened or which cannot be minimally opened are pruned and all
branches which can  be minimally opened are kept (and opened). (We
reconsider this pruning process in section
\ref{sec:implementation}.)

 According to Definition \ref{de:cons} and Theorem
 \ref{theo:repairop},
$\bar{t}$~ is a consistent answer to the open query $Q(\bar{x})$
when the combined tableau $\nit{op}(\TP(\IC \cup r)) \otimes
\TP(\neg Q(\bar{t}))$ (c.f. Definition \ref{def:branches}) is,
again, a closed tableau. In consequence, we might use the tableau
$\nit{op}(\TP(\IC \cup r)) \otimes \TP(\neg Q(\bar{x}))$ in order
to retrieve those values for $\bar{x}$ that restore the closure of
all the opened branches in the tableau.

\begin{example}\label{ex:stco}
Consider the functional dependency $$IC:~~
\forall(x,y,z,u,v)(Student (x,y,z)\wedge Student(x,u,v)
~\rightarrow~ y=u \wedge z=v);$$ and the inconsistent students
database instance \begin{eqnarray*} r &=& \{Student(S_1,N_1,D_1),
 Student(S_1,N_2,D_1),
 Course(S_1, C_1, G_1),\\
 &&
~~Course(S_1, C_2, G_2)\}, \end{eqnarray*} which has the two
repairs, namely $$r_1 = \{Student(S_1,N_1,D_1),  Course(S_1, C_1,
G_1), Course(S_1, C_2, G_2)\},$$  $$r_2 = \{ Student(S_1,N_2,D_1),
Course(S_1, C_1, G_1), Course(S_1, C_2, G_2)\}.$$

We can distinguish two kinds of queries. The first one corresponds
to a first order formula containing free variables (not
quantified), and then expects a (set of database) tuple(s) as
answer. For example, we want the consistent answers to
 the query ``$Course(x,y,z)?$".  Here
we have that $\nit{op}(\TP(\IC \cup r)) \otimes \TP(\neg
Course(x,y,z))$ is closed for the tuples $(S_1, C_1, G_1)$ and
$(S_1, C_2, G_2)$.

A second kind  of queries corresponds to  queries without free
variables, i.e. to sentences. They should get the answer ``yes''
or ``no''. For example, consider  the query ``$Course(S_1,$ $ C_2,
G_2)?$". Here $\nit{op}(\TP(\IC \cup r)) \otimes \TP($ $\neg
Course(S_1, C_2, G_2))$ is closed. The answer is ``yes'', meaning
that the sentence is true in all repairs.

Now, consider the query ``$Student(S_1,N_2,D_1)?$". The tableau
$\nit{op}(\TP(\IC \cup r)) \otimes \TP(\neg Student(S_1,N_2,D_1))$
is not closed, and $Student(S_1,N_2,D_1)$ is not a member of both
repairs. The answer is ``no'', meaning that the query is not true
in all repairs. \boxtheorem
\end{example}

The following example shows that, as opposed to
\cite{ArenasBertossiChomicki:99}, we are able to treat existential
queries in a proper  way.

\begin{example} \label{ex:exist}
Consider the query ``$\exists x Course(x, C_2, G_2)$?"  for the
database in example \ref{ex:stco}. Here we have that
$\nit{op}(\TP(\IC \cup r)) \otimes \TP(\neg \exists x ~Course(x,
C_2, G_2))$ is closed. The second tableau introduces the formulas
$\neg Course(p,C_2,G_2)$, for every $c \in D \cup P$ in every
branch. The answer is ``yes''. This answer has been obtained by
replacing  $p$ by the same constant $S_1$ in both branches. This
does not need to be always the case. For example, with the query
~``$\exists x ~Student(S_1,x,D_1)?$", that introduces the formulas
$\neg Student(S_1,p,D_1)$ in every branch of $\nit{op}(\TP(\IC
\cup r)) \otimes \TP(\neg \exists x ~Student(S_1,x,D_1))$, the
tableau closes, the answer is  ``yes'', but one repair has been
closed for $p =N_1$ and the other repair has been closed for $p =
N_2$.

We can also handle open existential queries. Consider now  the
query with $y$ as the free variable ~ ``$\exists z
Course(S_1,y,z)?$". The tableaux for $\nit{op}(\TP(\IC \cup r))
\otimes \TP(\neg \exists z Course(S_1,y,z))$, which introduces the
formulas $\neg Course(S_1,y,p)$ in every branch, is closed,
actually by $y = C_1$, and also by $y= C_2$, but for two different
values for $p$, namely $G_1$ and $G_2$, resp. \boxtheorem

\end{example}

\begin{theorem}\em
Let $r$ be an inconsistent database wrt to the set of integrity
constraints $IC$.
\be
\item Let $Q(\bar{x})$ be an open query with the free variables
$\bar{x}$. A ground tuple $\bar{t}$  is a consistent answer to
$Q(\bar{x})$ iff $\nit{op}(\TP(\IC \cup r)) \otimes \TP(\neg
Q(\bar{x}))$ is closed for the substitution ~$\bar{x} \mapsto
\bar{t}$.
\item Let $Q$ be query without free variables. The answer is ``yes'', meaning
that the query is true in all repairs, iff $\nit{op}(\TP(\IC \cup
r)) \otimes \TP(\neg Q)$ is closed. \ee
\end{theorem}

\section{CQA,  Minimal Entailment and Tableaux} \label{sec:circum}

As the following example shows, CQA is a form of {\em
non-monotonic entailment}, i.e. given a relational database
instance $r$, a set of ICs $\IC$, and a consistent answer
$P(\bar{a})$ wrt $\IC$, i.e. $r \models_c \varphi$, it may be the
case that $r^\prime  \not \models_c P(\bar{a})$, for an instance
$r^\prime$ that extends $r$.

\begin{example}\label{ex:nonmon} The database containing the
table
\begin{center}
\begin{tabular}{c|cc}
${\it Employee}$ & ${\it Name}$ & ${\it Salary}$\\ \hline & ${\it
J.Page}$ & 5000\\ & ${\it V.Smith}$ & 3000\\ & ${\it M.Stowe}$ &
7000\\
\end{tabular}
\end{center}
is consistent wrt the FD $f_1: {\it Name} \rightarrow {\it
Salary}$. In consequence, the set of consistent answers to the
query $Q(x,y):~ \nit{Employee}(x,y)$ ~is~ $\{({\it J.Page},~
5000), ({\it V.Smith},$ $ ~3000),$ $({\it M.Stowe},~ 7000)\}$. If
we add the tuple $({\it J. Page},~ 8000)$ to the database, the set
of consistent answers to the same query is reduced to $\{({\it
V.Smith}, ~3000),$ $({\it M.Stowe},$ $7000)\}$. \boxtheorem\\
\end{example}

We may be interested in having a logical specification
 ${\it
Spec_r}$ of the repairs of the database instance $r$.  In this
case,  we could consistently answer a query $Q(\bar{x})$, by
asking for those $\bar{t}$ such that
\begin{equation}\label{eq:nonmon}
{\it Spec}_r~ |\!\!\!\approx~ Q(\bar{t}) ~~~\equiv~~~ r\models_c
Q(\bar{t}),
\end{equation}
where $|\!\!\!\approx$ is a new, suitable consequence relation,
that, as the example shows,  has to be non-monotonic.

\subsection{A circumscriptive characterization of CQA}
\label{sec:circsp}
 Notice that with CQA we have a minimal
entailment relation in the sense that consistent answer are true
of certain minimal models, those that minimally differ from the
original instance. This is a more general reason for obtaining a
nonmonotonic consequence relation. Actually, the database repairs
can be specified by means of a circumscription axiom
\cite{mccarthy86,lifschitzSrvy} that has the effect of minimizing
the set of changes to the original database performed in order to
satisfy the ICs.

Let $P_1, \ldots, P_n$ be the database predicates in ${\cal L}$.
In the original instance $r$, each $P_i$ has a finite extension
that we also denote by $P_i$. Let $R_1, \ldots, R_n$ be new copies
of $P_1, \ldots, P_i$, standing for the corresponding tables in
the database repairs. Define, for $i = 1, \ldots, n$,
\begin{equation}
\forall \bar{x}[P_i^{\nit{in}}(\bar{x})
~~_{\nit{def}}\!\!\longleftrightarrow~~ (R_i(\bar{x}) \wedge \neg
P_i(\bar{x}))], \label{eq:difin}
\end{equation}
\begin{equation}
\forall \bar{x}[P_i^{\nit{out}}(\bar{x})
~~_{\nit{def}}\!\!\longleftrightarrow~~ (P_i(\bar{x}) \wedge \neg
R_i(\bar{x}))]. \label{eq:difout}
\end{equation}
Consider now the theory $\Sigma$ consisting of axioms
(\ref{eq:difin}), (\ref{eq:difout}) plus $r$, i.e. the (finite)
conjunction of the atoms in the database, plus $\IC(P_1/R_1,
\cdots, P_n/R_n)$, i.e. the set of ICs, but with the original
database predicates replaced by the new predicates; and possibly,
axioms for the built-in predicates, e.g. equality.

In order to minimize the set of changes, we circumscribe in
parallel the predicates $P_i^{\nit{in}}, P_i^{\nit{out}}$ in the
theory $\Sigma$, with variable predicates $R_1,\ldots, R_n$, and
fixed predicates $P_1, \ldots, P_n$ \cite{lifschitz}, that is, we
consider the following circumscription
\begin{equation}
\nit{Circum}(\Sigma; P_1^{\nit{in}}, \ldots P_n^{\nit{out}}; R_1,
\ldots, R_n; P_1, \ldots, P_n). \label{eq:circ}
\end{equation}
The semi-colons separate the theory, the predicates minimized in
parallel, the variable predicates and the fixed predicate, in that
order.

We want to minimize the  differences between a database repair and
the original database instance. For this reason we need the $R_i$
to be flexible  in the minimization process. The original
predicates $P_i$s are not subject to changes, because the changes
can be read  from the $R_i$ (or from their differences with the
$P_i$).

\begin{example}\label{ex:circum}
Consider $r = \{P(a)\}$ and $\IC = \{\forall x(P(x) \rightarrow
Q(x))\}$. In this case, $\Sigma$ consists of the following
sentences: ~$P(a), \forall x (R_P(x) \rightarrow R_Q(x)),$
$\forall x(P^{in}(x) \leftrightarrow R_P(x) \wedge \neg P(x)),$
$\forall x(P^{out}(x) \leftrightarrow P(x) \wedge \neg R_P(x)),$
$\forall x(Q^{in}(x) \leftrightarrow R_Q(x) \wedge \neg Q(x)),$
$\forall x(Q^{out}(x) \leftrightarrow Q(x) \wedge \neg R_Q(x))$.
Here the new database predicates are $R_P$ and $R_Q$. They vary
when $P^{in}, P^{out}, Q^{in}, Q^{out}$ are minimized.

The models of th circumscription are the minimal (classical)
models of the theory $\Sigma$. A model $\frak M = <M, (P^{in})^M,
(P^{out})^M, (Q^{in})^M, (Q^{out})^M, R_P^M, R_Q^M, P^M,$ $ Q^M,
a^M>$ is minimal if there is no other model with the same domain
$M$ that interprets $P, Q, a$ in the same way as $\frak M$ and has
at least one of the interpretations of $P^{in}$, $P^{out}$,
$Q^{in}$, $Q^{out}$ strictly included in the corresponding in
$\frak M$ and the others (not necessarily strictly) included in
the corresponding in $\frak M$. \boxtheorem
\end{example}

Circumscription (\ref{eq:circ}) can be specified by means of a
second-order axiom
\begin{eqnarray}\label{eq:axiom}
&\Sigma(P_1^{\nit{in}}, \ldots, P_n^{\nit{in}}, P_1^{\nit{out}},
\ldots, P_n^{\nit{out}}, R_1, \ldots, R_n) ~\wedge\\ &\forall X_1
\cdots \forall X_n \forall Y_1 \cdots \forall Y_n \forall Z_1
\cdots \forall Z_n(\Sigma(X_1, \ldots, X_n, Y_1, \ldots, Y_n, Z_1,
\ldots, Z_n) ~\wedge \nonumber\\
 & \bigwedge_1^n X_i \subseteq P_i^{\nit{in}}
~\wedge~ \bigwedge_1^n Y_i \subseteq P_i^{\nit{out}}
~\longrightarrow~ \bigwedge_1^n P_i^{\nit{in}} \subseteq X_i
~\wedge~ \bigwedge_1^n  P_i^{\nit{out}} \subseteq Y_i). \nonumber
\end{eqnarray}
The first conjunct emphasizes the fact that the theory is
expressed in terms of the predicates shown there. Those predicates
are replaced by second-order variables in the  $\Sigma$ in the
quantified part of the formula. The circumscription axiom says
that the change predicates $R_i^{\nit{in}}, R_i^{\nit{out}}$ have
the minimal extension under set inclusion among those that satisfy
the ICs. It is straightforward to prove that the database repairs
are in one to one correspondence with the restrictions to $R_1,
\ldots, R_n$ of those Herbrand models  of the circumscription that
have domain $D$ and the  extensions of the predicates $P_1,
\ldots, P_n$ as  in the original instance $r$.

An alternative to externally fixing the domain $D$ consists in
minimizing the finite active domain, that is a subset of $D$. This
can be achieved by means of a circumscription as well, and then
that domain can be extended to the whole of $D$. Notice that in
order to capture the unique names assumption of databases, the
equality predicate could be minimized. Furthermore, if we want the
minimal models to have the extensions for the $P_i$ as in $r$, we
can either include in $\Sigma$ predicate closure axioms of the
form $\forall \bar{x} (P_i(\bar{x}) \leftrightarrow
\bigvee_1^{k_i} \bar{x}_j = \bar{a}_j)$ if $P_i$'s extension is
non-empty and $\forall \bar{x} (P_i(\bar{x})) \leftrightarrow
\bar{x}\neq \bar{x})$ if it is empty; or apply to those predicates
the closed world assumption, that can also be captured by means of
circumscription. See \cite{lifschitzSrvy} for details. Another
alternative is to fix the domain $D$ and replace everywhere $r$ in
$\Sigma$ by the first-order sentence, $\sigma(r)$, corresponding
to Reiter's logical reconstruction of database instance $r$
\cite{reiter:84}. We do not do any of this explicitly, but leave
it as something to be captured at the implementation level.

\begin{example}\label{ex:circ2} (example \ref{ex:circum}
continued) The minimal model of the circumscription of the theory
are~ $<D,\emptyset,\{a\}, \emptyset, \emptyset, \emptyset,
\emptyset, \{a\}, \emptyset>$ and $<D, \emptyset, \emptyset,
\{a\}, \emptyset, \{a\}, \{a\},$ $ \{a\}, \emptyset>$, that show
first the domain and next the extensions of $P^{in}, P^{out},
Q^{in},$ $ Q^{out}, R_P, R_Q, P, Q>$, in this order. The first
model corresponds to repairing the database by deleting $P(a)$;
the second, to inserting $Q(a)$. \boxtheorem
\end{example}

By playing with different kinds of circumscription, e.g.
introducing priorities \cite{lifschitz}, or considering only some
change predicates, e.g. only $P_i^{out}$'s (only deletions),
preferences for some particular kinds of database repairs could be
captured. We do not explore here this direction any further.

The original theory $\Sigma$ can be written as $\Sigma^\prime
\wedge r$, where $\Sigma^\prime$ is formed by all the conjunctions
in $\Sigma$, except for $r$. It is easy to see that the
circumscription $\nit{Circum}(\Sigma; R_1^{\nit{in}}, \ldots
R_n^{\nit{out}}; R_1, \ldots, R_n; P_1, \ldots, P_n)$ is logically
equivalent to $r \wedge  \nit{Circum}(\Sigma^\prime;
R_1^{\nit{in}}, \ldots R_n^{\nit{out}}; R_1, \ldots, R_n; P_1,
\ldots, P_n)$. In consequence, we can replace (\ref{eq:nonmon}) by
\begin{equation}\label{eq:nonmon2}
r \wedge \nit{Circum}(\Sigma^\prime; R_1^{\nit{in}}, \ldots
R_n^{\nit{out}}; R_1, \ldots, R_n; P_1, \ldots, P_n) \models
Q(\bar{t}) ~~~\equiv~~~ r\models_c Q(\bar{t}).
\end{equation}
We can see that in this case the nonmonotonic consequence relation
$|\!\!\!\approx$ corresponds then to classical logical
consequence, but with the original data put in conjunction with a
second-order theory.

Some work has been done on detecting conditions and developing
algorithms for the collapse of a (second-order) circumscription to
a first-order theory \cite{lifschitz,doherty}. The same for
collapsing circumscription to logic programs \cite{gl89}. In our
case, this would not be surprising. In \cite{fqas2k,greco3,nmr02},
direct specifications of database repairs by means of logic
programs are presented.

In our case, there is not much hope in having the circumscription
collapse to a first-order sentence, $\varphi_{\nit{Circ}}$. If
this were the case, CQA would be feasible in polynomial time in
the size of the database, because then for a query  $Q$, the query
$(\varphi_{\nit{Circ}} \rightarrow Q)$ could be posed to the
original instance $r$. As shown in \cite{janIPL}, CQA can be
coNP-complete, even with simple functional dependencies and
(existentially quantified) conjunctive queries. Actually, in the
general case CQA is indecidable  (to appear in an extended version
of \cite{ArenasBertossiChomicki:99}).

Under those circumstances, it seems a natural idea to explore to
what extent semantic tableaux can be used for CQA. Actually, some
implementations to nomonotonic reasoning, more precisely to
minimal entailment, based on semantic tableaux have been proposed
in \cite{olivetti92,niemela96a,niemela96b,olivetti99,bry2k}.

\subsection{Towards implementation} \label{sec:implementation}

The most interesting proposal for implementing first order
circumscriptive reasoning with semantic tableaux is offered by
Niemela in \cite{niemela96b}, where optimized techniques for
developing tableaux branches and checking their minimality are
introduced. The techniques presented there, that allow minimized,
variable and fixed predicates, could be applied in our context,
either directly, appealing to the circumscriptive characterization
of CQA we gave before, or adapting Niemela's techniques to the
particular kind of process we have at hand, in terms of minimal
opening of branches in the tableau $\TP(\IC \cup
r)$.\footnote{Notice that the input theory in this case differs
from the theory to which the circumscription is applied in the
previous section.} We will briefly explore this second
alternative.

As in \cite{niemela96b}, we assume in this section that (a) the
semantic tableaux are applied to formulas in clausal form, and (b)
only  Herbrand models are considered, what in our case represents
no limitation, because our openings, repairs, etc. are all
Herbrand structures. Furthermore, if $\IC$ contains {\em safe}
formulas \cite{ullmanI}, what is commonly required in database
applications, we can restrict the Herbrand domain to be the finite
active domain of the database.

As seen in section \ref{se:QueryAnswering}, consistently answering
query $Q$ from instance $r$ wrt $\IC$, can be based on the
combination of  $\nit{op}(\TP(\IC \cup r))$ and $\TP(\neg
Q(\bar{x})$. Nevertheless, explicitly having the first, pruned,
tableau  amounts to having also explicitly all possible repairs of
the original database. Moreover, this requires having verified the
property of minimality in the data closed branches, possibly
comparing different branches wrt to inclusion. It is more
appealing to check minimality as the tableau $\TP(\IC \cup r)$ is
developed.

Notice that if a finished branch $B \in \TP(\IC \cup r)$, opened
after a preliminary data closure was reached, remains open for
$\bar{x} = \bar{t}$ when combined  with $\TP(\neg Q(\bar{x}))$,
then  $\opb$ is a model of $\IC$ and $\neg Q(\bar{t})$, and in
consequence $\opb$ provides a counterexample to $\IC \models
Q(\bar{t})$. However, this is classical entailment, and we are
interested in those models of $\IC$ that minimally differ from
$r$, in consequence, $\opb$ may not be a counterexample for our
problem of CQA, because the it may not correspond to a repair of
the original instance. Such branches that would lead to a non
minimal opening in $\TP(\IC \cup r)$ should be closed, and left
closed exactly as those branches that were closed due to
built-ins.

As we can see, what is needed is a methodology for developing the
tableaux such that: (a) Each potential counterexample is explored,
and hopefully at most once. (b) Being a non minimal opening is
treated as a closure condition (because, as we just saw,  they do
not provide appropriate counterexamples). (c) The minimality
condition is checked locally, without comparison with other
branches, what is much more efficient in terms of space.

Such methodology is proposed in \cite{niemela96b}, with two
classical rules for generating tableaux, a kind of hyper-type
rule, and a kind of cut rule.  The closure conditions are as in
the classical case, but a new closure condition is added, to close
branches that do not lead to minimal models. This is achieved by
means of a ``local" minimality test, that can also be found in
\cite{niemela96a,eiter93}. We can
 adapt and adopt such a test in our framework on the basis of the definition of
 {\em grounded} model given in \cite{niemela96b} and our circumscriptive
 characterization of CQA given above.

  Let $B$ be a data closed branch in $\TP(\IC \cup r)$, with
  $\nit{op}(B) =  (r \setminus L) \cup K$. We associate to
  $B$ a Herbrand structure $M(B)$ over the first order language
  ${\cal L}(\bar{K}, \bar{L}, \bar{P}, \bar{R})$, where $\bar{R} = <R_1, \ldots, R_n>$ is the
  list of original database predicates, $\bar{P} = <P_1, \ldots, P_n>$
  is the list of predicates for the repaired versions of the
  $R_i$s, $\bar{L} = <L_1, \ldots, L_n>, \bar{K} = <K_1, \ldots, K_n>$ are
  predicates for $R_i \setminus P_i$ and $P_i \setminus R_i$,
  resp. (Then it makes sense to identify the list of predicates
  $\bar{L}$ and $\bar{K}$ with the sets of differences $K$ and $L$ in the
  branch $B$). $M(B) = <\nit{Act}(r), \bar{L}^B, \bar{K}^B, \bar{P}^B, \bar{R}^B>$ is defined through (and can be identified
  with) the subset
  $\Lambda := \bigcup_1^n L_i^B \cup \bigcup_1^n K_i^B \cup \bigcup_1^n P_i^B
  \cup \bigcup_i^n R_i^B$ of the Herbrand base ${\cal B}$, where  $\bigcup_1^n R_i^B$
  coincides with the database contents  $r$, and the elements in
  $\bigcup_1^n P_i^B$ are taken from $\opb$.

  Now we can reformulate for our context the notion of grounded Herbrand structure given in
  \cite{niemela96b}.

  \begin{definition} \label{def:ungrounded} \em (adapted from \cite{niemela96b}) An opening
  $\opb$ is {\em grounded} iff
for all $p \in \bar{K} \cup \bar{L}$ with
  $p(\bar{t}) \in \Lambda$ it holds

  \vspace{-5mm}
\begin{equation}
\IC(P_1/R_1, \ldots, P_n/R_n) ~\cup~ \{\bigwedge_i^n (L_i = R_i
\setminus
  P_i), \bigwedge_i^n K_i = P_i \setminus R_i)\}
  \label{eq:theory}
  \end{equation}

  \vspace{-3mm}\hspace*{7.8cm}$\cup~ N^{<\bar{L},\bar{K};\bar{R}>}(\Lambda)
   ~\models~ p(\bar{t}),$

  \noindent where~
  $N^{<\bar{L},\bar{K};\bar{R}>}(\Lambda) := \{\neg q(\bar{t}) ~|~
  q \in \bar{L} \cup \bar{K} \cup \bar{R} \mbox{ and } q(\bar{t})
  \in {\cal B} \setminus \Lambda\} ~\cup$ \\
  \hspace*{3.7cm} $\{q(\bar{t}) ~|~ q \in
  \bar{R} \mbox{ and } q(\bar{t}) \in \Lambda\}.$ \boxtheorem
\end{definition}

Notice that the first set in the union that defines
$N^{<\bar{L},\bar{K};\bar{R}>}(\Lambda)$ corresponds to the CWA
applied to the minimized predicates. i.e. those in $\bar{L},
\bar{K}$, and the fixed predicates, i.e. those in $\bar{R}$. The
second set coincides with the original database contents $r$. From
the results in \cite{niemela96b} and  our circumscriptive
characterization of CQA, we obtain the following theorem.

\begin{theorem}\em An opening $\opb$ corresponds to a  database repair
iff $M(B)$ is a grounded model of (\ref{eq:theory}). \boxtheorem
\end{theorem}

Ungrounded models can be discarded, and then ungroundedness can be
used as an additional closure condition on branches. Notice that
the test is local to a branch and can be applied at any stage of
the development of a branch, even when it is not finished yet. The
test is based on classical logical consequence, and then not on
any kind of minimal entailment.

\begin{example} (example \ref{ex:models2} continued)
We need some extra predicates. $P_P, P_Q, P_R$ stand for the
repaired versions of $P, Q, R$, resp. $L_P, L_Q, L_R, K_P, K_Q,
K_R$ stand for $P \setminus P_P, \ldots, P_R \setminus R$, resp.
Here $\bar{L} = <L_P,L_Q,L_R>, \bar{K} = <K_P,K_Q,K_R>, \bar{P} =
<P_P, P_Q, P_R>, \bar{R} = <P,Q,R>$.

In order to check groundedness for branches, we have the
underlying theory ~$\Sigma = \{\forall x(P_P(x) \rightarrow
P_Q(x)), \forall x (L_P(x) \leftrightarrow (P(x) \wedge \neg
P_P(x))), \ldots, \forall x(K_R(x) \leftrightarrow (P_R(x) \wedge
\neg R(x)))\}$, corresponding to (\ref{eq:theory}).

In order to check the minimality of branch $B_1$, we consider
$M(B_1)$, that is determined by the set of ground atoms
$\Lambda(B_1) = \{P(a), R(b), L_P(a),  R_R(b)\}$. First, this
structure satisfies $\Sigma$. Now, for this branch
\begin{eqnarray*}
N^{<\bar{L},\bar{K};\bar{R}>}(\Lambda(B_1)) &=& \{\neg L_P(b),
\neg L_Q(a), \neg L_Q(b), \neg L_R(a), \neg L_R(b), \neg K_P(a),\\
&& ~~\neg K_P(b), \neg K_Q(a), \neg K_Q(b), \neg K_R(a), \neg
K_R(b), \neg P(b),\\&& ~~\neg Q(a), \neg Q(b), \neg R(a)\} ~\cup~
\{P(a), R(b)\}.
\end{eqnarray*}
For groundedness, we have to check if $L_P(a)$ is a classical
logical consequence of $\Sigma \cup
N^{<\bar{L},\bar{K};\bar{R}>}(\Lambda(B_1))$. This is true,
because, from $\neg K_Q(a)$, we obtain $\neg P_Q(a)$. Using the
contrapositive of the IC in $\Sigma$, we obtain, $\neg P_P(a)$.

In consequence, the opening corresponding to branch $B_1$ is a
repair of the original database.

Consider now the unfinished branch $B_3$, for which $\Lambda(B_3)
= \{P(a), R(b),$ $K_Q(b),  R_P(a), R_Q(b), R_R(b)\}$, and
\begin{eqnarray*}
N^{<\bar{L},\bar{K};\bar{R}>}(\Lambda(B_3)) &=& \{\neg L_P(a),
\neg L_P(b), \neg L_Q(a), \neg L_Q(b), \neg L_R(a), \neg L_R(b),\\
&& ~~\neg K_P(a), \neg K_P(b), \neg K_Q(a), \neg K_R(a), \neg
K_R(b), \neg P(b),\\&& ~~\neg Q(a), \neg Q(b), \neg R(a)\} ~\cup~
\{P(a), R(b)\}.
\end{eqnarray*}
\end{example}
We have to apply the groundedness test to $K_Q(b)$. In this case
it is not possible to derive this atom from $\Sigma \cup
N^{<\bar{L},\bar{K};\bar{R}>}(\Lambda(B_3))$, meaning that the set
of literal is not grounded. If we keep developing that branch, the
set $N$ can only shrink. In consequence, we will not derive the
atom in the extensions. We can stop developing  branch $B_3$
because we will not get a minimal opening. \boxtheorem

\section{Conclusions}\label{se:concl}

We have presented the theoretical basis for a treatment of
consistent query answering in relational databases by means of
analytic tableaux. We have mainly concentrated on the interaction
of the database instance and the integrity constraints; and in the
problem of representing  database repairs by means of opened
tableaux. However, we also showed how the analytic tableaux
methodology could we also used for consistent query answering.

We established the connections between the problem of consistent
query answering and knowledge base update, on one side, and
circumscriptive reasoning, on the other. This is not surprising,
since the relationship between knowledge base update and
circumscription has already been studied by Winslett
\cite{Winslett:89,Winslett:91} (see also \cite{liberatore}).

The connection of CQA to updates and minimal entailment allowed us
to apply know complexity results to our scenario.  Furthermore, we
have seen that the reformulation of the problem of CQA as one of
computing circumscription opens the possibility of applying
established methodologies for semantic tableaux based
methodologies for circumscriptive reasoning.

As we have seen, there are several similarities between our
approach to consistency handling  and those followed by the belief
revision/update community. Database repairs coincide with revised
models defined by Winslett in \cite{Winslett:88}. The treatment in
\cite{Winslett:88} is mainly propositional, but a preliminary
extension to first order knowledge bases can be found in
\cite{winslett94}. Those papers concentrate on the computation of
the models of the revised theory, i.e., the repairs in our case,
but not on query answering. Comparing our framework with that of
belief revision, we have an empty domain theory, one model: the
database instance, and a revision by a set of ICs. The revision of
a database instance by the ICs produces new database instances,
the repairs of the original database.

Nevertheless, our motivation and starting point are quite
different from those of belief revision. We are not interested in
computing the repairs {\em per se}, but in answering queries,
hopefully using the original database as much as possible,
possibly posing a modified query. If this is not possible, we look
for methodologies for representing and querying simultaneously and
implicitly all the repairs of the database. Furthermore, we work
in a fully first-order framework. Other connections to belief
revision/update can be found in  \cite{ArenasBertossiChomicki:99}.

To the best of our knowledge, the first treatment of CQA in
databases goes back to \cite{Bry97}. The approach is based on a
purely proof-theoretic notion of consistent query answer. This
notion, described only in the propositional case, is more
restricted than the one we used in this paper. In
\cite{Cholvy:98}, Cholvy presents a general logic framework for
reasoning about contradictory information which is based on an
axiomatization in modal propositional logic. Instead, our approach
is based on classical first order logic.

Other approaches to consistent query answering  based on logic
programs with stable model semantics were presented in
\cite{fqas2k,nmr02,greco3}. They can handle general first order
queries with universal ICs.

There are many open issues. One of them has to do with the
possibility of  obtaining from the tableaux for instances and ICs
the right ``residues" that can be used  to rewrite a query as in
\cite{ArenasBertossiChomicki:99}.  The theoretical basis of CQA
proposed in \cite{ArenasBertossiChomicki:99} were refined and
implemented in \cite{CelleBertossi:2K}. Comparisons of the
tableaux based methodology for CQA and the ``rewriting based
approach" presented in those papers is an open issue. However,
query rewriting can not be applied to existential queries like the
one in example \ref{ex:exist}, whereas the tableaux methodology
can be used. Perhaps, an appropriate use of tableaux could make
possible an extension of the rewriting approach to syntactically
richer queries and ICs.

Another interesting open issue  has to do with the fact that we
have treated Skolem parameters  as null values. It would be
interesting to study the applicability in our scenario of
methodologies for  query evaluation in databases in the presence
of  null values like the one presented in \cite{reiterNV}.

 In this paper we have concentrated mostly on the theoretical foundations
 of a methodology based on semantic tableaux for querying inconsistent databases.
 Nevertheless, the methodology for CQA  requires further investigation. In this context, the most
interesting open problems have to do with implementation issues.
More specifically, the main challenge consists in developing
heuristics and mechanisms for using a tableaux theorem prover to
generate/store/represent $\nit{TP}(\IC \cup r)$ in a compact form
with the purpose of: (a) applying the database assumptions, (b)
interacting with a DBMS on request, in particular, without
replicating the whole database instance at the tableau level, (c)
 detecting and producing the minimal openings (only), (d)
using a theorem prover (in combination with a DBMS) in order to
consistently answer  queries.

An important issue in database applications is that usually
queries have free variables and then answer sets have to be
retrieved as a result of the automated reasoning process. Notice
that once we have $\nit{op}(\nit{TP}(\IC \cup r))$, we need to be
able to: (a) use it for different queries ~$Q$, (b)
 process the combined tableau $\nit{op}(\nit{TP}(\IC \cup
r)) \otimes \nit{TP}(\neg Q)$ in an ``reasonable and practical"
way. We have seen that existing methodologies and algorithms like
the one presented in \cite{niemela96b}, can be used in this
direction. However, producing a working implementation,
considering all kinds of optimizations with respect to
representation and development of the tableaux, grounding
techniques, database/theorem-prover interaction, etc. is a major
task that deserves  separate investigation.

\vspace{2mm}\noindent {\bf Acknowledgments:} ~Work supported by
FONDECYT Grant \# 1000593;  ECOS /CONICYT Grant C97E05, Carleton
University Start-Up Grant 9364-01, and NSERC Grant  250279-02.
Preliminary versions of this paper appeared in
\cite{WS01,foiks02}; we are grateful to anonymous referees for
their remarks.

\bibliographystyle{plain}

\begin{thebibliography}{12}

\bibitem{ArenasBertossiChomicki:99}
Arenas, A.,  Bertossi, L. and Chomicki, J.
\newblock Consistent Query Answers in Inconsistent Databases.
\newblock Proc. ACM Symposium on Principles of Database Systems
(ACM PODS'99). ACM Press,  1999, pp. 68--79.

\bibitem{fqas2k}
Arenas, M.; Bertossi, L. and Chomicki, J.
\newblock {Specifying and Querying Database Repairs using Logic Programs with
  Exceptions}.
\newblock In
{\em Flexible Query Answering Systems. Recent Developments}, H.L.
Larsen, J. Kacprzyk, S. Zadrozny, H. Christiansen (eds.).
Springer-Verlag, 2000, pp. 27--41.

\bibitem{scalar}
Arenas, A.,  Bertossi, L. and Chomicki, J.
\newblock Scalar Aggregation in FD-Inconsistent Databases.
\newblock In Database Theory - ICDT 2001 (Proc. International Conference
on Database Theory, ICDT'2001). Springer LNCS 1973, 2001, pp. 39
-- 53.

\bibitem{WS01}
Bertossi, L. and Schwind, C. B.
\newblock An Analytic Tableaux based Characterization of Database Repairs
for Consistent Query Answering (preliminary report).
\newblock In Working Notes of the
IJCAI'01 Workshop on Inconsistency in Data and Knowledge. AAAI
Press, 2001, pp. 95 -- 106.

\bibitem{foiks02}
Bertossi, L. and Schwind, C. B.
\newblock{Analytic Tableaux and Database Repairs: Foundations}.
\newblock In {\em Foundations of Information and Knowledge Systems (Proc. FoIKS 2002)}, Eiter, T.
and Schewe, K.-D. (eds.). Springer LNCS 2284, 2002, pp. 32-48.

\bibitem{nmr02}
Barcelo, P. and Bertossi, L.
\newblock {Repairing Databases with Annotated Predicate Logic}.
In {\em Proc. Ninth International Workshop on Non-Monotonic
Reasoning (NMR'2002). Special session on Changing and Integrating
Information: From Theory to Practice}. S. Benferhat and E.
Giunchiglia (eds.). Morgan Kaufmann Publishers, 2002, pp. 160 --
170.

\bibitem{ABelletal:95}
Belleann\'ee, C.,  Kuhna, P.,
  Lamarre, P.,  Schwind, C.,
  Thi\'ebaux, S.,  Vialard, V.  and Vorc'h, R.
\newblock  M\'{e}thodes
S\'{e}mantiques de D\'{e}monstration pour Logiques Non-standards.
\newblock In {\em
PRC GDR Intelligence artificielle, Actes des 5\`{e}mes
Journ\'{e}es Nationales}, Nancy 2-5, f\'{e}vrier 1995.


\bibitem{Beth:59}
Beth, E. W.
\newblock {\em The Foundations of Mathematics}.
\newblock North Holland, 1959.

\bibitem{Bry97}
 Bry, F.
\newblock Query Answering in Information Systems with Integrity
Constraints.
\newblock In  Proc. IFIP WG 11.5 Working Conference on Integrity and Control in
    Information Systems,
Chapman \& Hall, 1997.

\bibitem{bry2k}
Bry, F. and Yahya, A.H. \newblock {Positive Unit Hyperresolution
Tableaux and Their
       Application to Minimal Model Generation}.
       \newblock {\em Journal of Automated Reasoning},  25(1) (2000)
       35--82.



\bibitem{CelleBertossi:2K}
Celle, A. and Bertossi, L.
\newblock Querying Inconsistent Databases: Algorithms and
Implementation.
\newblock In `Computational Logic - CL 2000', J. Lloyd et al.
(eds.). Stream: 6th International Conference on Rules and Objects
in Databases (DOOD'2000). Springer LNAI 1861,  2000, pp. 942 --
956.

\bibitem{Cholvy:98}
Cholvy, L.
\newblock A General Framework for Reasoning about Contradictory
Information  and some of its Applications.
\newblock In Proceedings of ECAI Workshop ``Conflicts among Agents", Brighton, England,
August 1998.

\bibitem{janIPL}
Chomicki, J. and Marcinkowski, J.
\newblock {On the Computational Complexity of Consistent Query Answers}.
\newblock Submitted in 2002 (CoRR paper cs.DB/0204010).


\bibitem{winslett94}
Chou, T. and Winslett, M.
\newblock {A Model-Based Belief Revision System}.
\newblock {\em J. Automated Reasoning}, 12 (1994) 157--208.


\bibitem{doherty}
Doherty, P., Lukaszewicz, W. and Szalas, A. \newblock {Computing
Circumscription
       Revisited: A Reduction Algorithm}.
       \newblock {\em Journal of Automated Reasoning},  18(3)
       (1997)
       297--336.

       \bibitem{EiterGottlob:92}
Eiter, T. and Gottlob, G.
\newblock On the Complexity of Propositional Knowledge Base Revision, Updates, and Counterfactuals.
\newblock {\em Artificial Intelligence},  57 (1992) 227-270.

\bibitem{eiter93}
Eiter, T. and Gottlob, G.
\newblock {Propositional Circumscription and Extended Closed World
Assumption are $\Pi^p_2$-complete}. \newblock {\em Theoretical
Computer Science}, 114 (1993) 231-245.

\bibitem{FittingJAR:88}
Fitting, M.
\newblock First Order Modal Tableaux.
\newblock {\em Journal of Automated Reasoning}, 4(2) (1988) 191--213.

\bibitem{Fitting96}
Fitting, M.
\newblock {\em First Order Logic and Automated Theorem Proving}.
\newblock Texts and Monographs in Computer Science. Springer-Verlag, 2nd Edition, 1996.

\bibitem{gl89}
Gelfond, G. and Lifschitz, V.
\newblock {Compiling Circumscriptive Theories into Logic Programs}.
\newblock In {\em Non--Monotonic Reasoning}. Springer LNAI 346,
  1989, pp. 74--99.

  \bibitem{Gottlob:92}
Gottlob, G.
\newblock Complexity Results for Nonmonotonic Logics.
\newblock {\em Journal of Logic and Computation}, 2(3) (1992).

  \bibitem{greco3}
Greco, G.; Greco, S. and Zumpano, E.
\newblock {A Logic Programming Approach to the Integration,
Repairing and Querying of  Inconsistent Databases}.
\newblock In  {\em Proc. 17th International
Conference on Logic Programming (ICLP'01)}, Ph. Codognet (ed.).
Springer LNCS 2237, 2001, pp. 348--364.

\bibitem{LafonSchwind:88}
Lafon, E. and Schwind, C. B.
\newblock A Theorem Prover for Action Performance.
\newblock In Y.~Kodratoff, editor, {\em Proceedings of the 8th European
  Conference on Artificial Intelligence},  Pitman Publishing,
  1988, pp. 541--546.


\bibitem{liberatore}
Liberatore, P. and Schaerf, M.
\newblock {Reducing Belief Revision to Circumscription (and vice versa)}.
\newblock {\em Artificial Intelligence}, 93 (1997) 261--296.


\bibitem{lifschitz}
Lifschitz, V.
\newblock {Computing Cirscumscription}.
\newblock In Proc. IJCAI'85, 1985, pp. 121-127.

\bibitem{lifschitzSrvy}
Lifschitz, V.
\newblock {Circumscription}.
\newblock In Handbook of Logic in Artificial Intelligence and
Logic Programming, Vol. 3. Oxford University Press, 1994, pp.
297--352.

\bibitem{lloyd87}
Lloyd, J.W.
\newblock {\em {F}oundations of {L}ogic {P}rogramming}.
\newblock Springer-Verlag, 1987.

\bibitem{mccarthy86}
McCarthy, J.
\newblock {Applications of Circumscription to Formalizing Common Sense
Knowledge}.
\newblock {\em Artificial Intelligence}, 26(3) (1986) 89--118.


\bibitem{niemela96a}
Niemela, I. \newblock {A Tableau Calculus for Minimal Model
Reasoning}. \newblock In Proc.  Fifth Workshop on Theorem Proving
with Analytic Tableaux and Related Methods. Springer LNCS 1071,
1996, pp. 278-294.

       \bibitem{niemela96b}
Niemela, I. \newblock {Implementing Circumscription Using a
Tableau Method}. \newblock Proc. ECAI 1996, pp. 80--84.

\bibitem{olivetti92}
Olivetti, N. \newblock {Tableaux and Sequent Calculus for Minimal
Entailment}. \newblock {\em Journal of
       Automated Reasoning}, 9(1) (1992) 99--139.

       \bibitem{olivetti99}
Olivetti, N. \newblock {Tableaux for Nonmonotonic Logics}.
\newblock In Handbook of Tableaux Methods. Kluwer Publishers, 1999, pp.
469--528.

\bibitem{reiter:84}
Reiter, R.
\newblock  Towards a Logical Reconstruction of Relational
                 Database Theory.
\newblock In `On Conceptual Modeling', Brodie, M.~L. and Mylopoulos, J. and Schmidt,
J.~W. (eds.). Springer-Verlag, 1984, pp. 191--233.

\bibitem{reiterNV}
Reiter, R.
\newblock
A Sound and Sometimes Complete Query Evaluation Algorithm for
               Relational Databases with Null Values.
               \newblock {\em Journal of the ACM}, 33(2) (1986) 349--370.


\bibitem{Schwind:90}
Schwind, C. B.
\newblock A Tableau-based Theorem Prover for a Decidable Subset of Default
  Logic.
\newblock In M.~E. Stickel, editor, {\em Proceedings of the 10th International
  Conference on Automated Deduction}.  Springer LNAI 449, 1990, pp. 541--546.

\bibitem{RischSchwind:94}
Schwind, C. B. and Risch, V.
\newblock Tableau-based Characterisation and Theorem Proving for Default Logic.
\newblock {\em Journal of Automated Reasoning}, 13(4)
(1994) 223--242.

\bibitem{Smullyan:68}
Smullyan, R. M.
\newblock {\em First Order Logic}.
\newblock Springer-Verlag, 1968.

\bibitem{ullmanI}
 Ullman, J.
\newblock {\em Principles of {D}atabase and {K}nowledge-{B}ase Systems, {V}ol.
{I}}.
\newblock Computer Science Press, 1988.

\bibitem{Winslett:88}
Winslett, M.
\newblock Reasoning about Action with a Possible Models Approach.
\newblock In {\em Proceedings of the 8th National Conference on Artificial
  Intelligence}, 1988, pp. 89--93.


\bibitem{Winslett:91}
Winslett, M.
\newblock {Cirscumscriptive Semantics for Updating Knowledge Bases}.
\newblock {\em Annals of Mathematics and Artificial Intelligence}, 3(2-4)
(1991) 429--.

\bibitem{Winslett:89}
Winslett, M.
\newblock {Sometimes Updates are Circumscription}.
\newblock {\em Proceeding of the International Joint Conference on Artificial Intelligence (IJCAI'89)}, 1989, pp. 859-863.



\end{thebibliography}

\section*{Appendix: Proofs}

\subsection*{Proof of Lemma   \ref{le:openings}}
%\begin{proof}
We have by Lemma \ref{le:disord} $r'\Delta r = r \setminus r'$ and
$r''\Delta r = r \setminus r''$. Then $l\in r'\Delta r$ iff $l\in
r \setminus r'$, i.e. $l\in r $ and $l\not\in r'$ from which it
follows that $l\in r $ and $l\not\in r''$. Hence $l\in r\setminus
r'' = r''\Delta r$.
%\end{proof}

\subsection*{Proof of Lemma    \ref{le:LK}}
%\begin{proof}
Let $r'$ be an opening of $r$. Then $r' = (r\setminus L) \cup K$,
where $L = \{l : l\in r {\mbox{ and }} \neg l \in I\}$ and $K =
\{l : l\in I {\mbox{ and there is no substitution }}  \sigma
{\mbox{ such that }} l\sigma \in r\}$. Let us first observe that
$L\cap K = \emptyset$ since $ L \subseteq r$ and for $l\in K$,
$l\not\in r$. We show that $r\Delta r' = L\cup K$. Let be $x\in
r\Delta r'$. \\ 1. Case  $x\in r $ and $x\not\in r'$. Then
$x\not\in K$ and $x \not\in (r\setminus L)$. But from this, we get
$x\in L$, hence $x\in L\cup K$\\ 2. Case $x\not\in r $ and $x\in
r'$, iff $x\not\in r $ and (($x \in r$ and $x\not\in L$) or $x\in
K$), iff $x\not\in r $ or $x\in K$ from which it follows  $x\in
K\cup L$.

On the other hand, let be $x\in L\cup K$. Again, we consider two
cases:\\ 1. Case $x\in L$, then by definition, $\neg x \in I$.
Then, $x\not\in r\setminus L$ and, since $I$ is open, $x\not\in
I$. From this, we get $x\not\in K$ and, since $r' = (r\setminus
L)\cup K$, $   x \not\in r'$, from which it follows that $x\in
r\Delta r'$.\\ 2. Case $x\in K$, then $x\in I$ and $x\not\in r$.
But then $x\in r'$ and therefore $x\in r\Delta r'$.
%\end{proof}

\subsection*{Proof of Proposition    \ref{theo:leqr}}
%\begin{proof}
By Lemma \ref{le:LK}, we have $r\Delta r_1 = L_1\cup K_1$ and
$r\Delta r_2 = L_2\cup K_2$. From $r_1 \leq_r r_2$ we get then
$L_1\cup K_1 \subseteq L_2\cup K_2$. Since $L_i \cap K_i =
\emptyset$, we have $L_1 \subseteq L_2$ and $K_1 \subseteq K_2$.
%\end{proof}

\subsection*{Proof of Theorem    \ref{theo:repairop}}
%\begin{proof}
Let $r'$ be a repair of $r$. Then $r' \models IC$ and $r' \in
Min_{leq_r}(ic)$.  Since $r'$ is a model of $IC$, by Theorem
\ref{theo:modtab}, $r'$ contains an open branch $I$ of the tableau
$\TP(\IC)$ for $\IC$. We have $r' = (r\setminus L) \cup K$ and
since $r'$ is minimal wrt  $\leq_r$, there is no $r''$ closer to
$r$ than $r'$.  i.e. there is no $r'' = (r\setminus L') \cup K'$
such that $L'\subset L$ and $K'\subset K$. Hence $r' \cup I$ is a
minimal opening  of $r\cup I$.

On the other hand, let $I\cup r'$ be a minimal opening of $I\cup
r$ in $\TP(\IC \cup r)$ where $I$ is an open branch of $\TP(\IC)$.
Then, by Definition \ref{de:opening}, $ r' = (r\setminus L) \cup
K$ where $L = \{l: l\in r$ and $\neg l \in I$ and $K= \{l: l\in I
$ and there is no substitution $\sigma$ such that $l\sigma \in
r\}$. By Lemma \ref{le:LK}, we have $r\Delta r'= L\cup K$. Since
$I\cup r$ is a minimal opening of $I\cup r'$, we have by Theorem
\ref{theo:leqr}, that there is no $r''$, $L''$ and $K''$ such that
$r''$ is an opening of $r$ and  $ r'' = (r\setminus L'') \cup K''$
and $L''\subset L$ and $K''\subset K$. By Lemma \ref{le:LK}, this
means that there is no $r''$ such that $r\Delta r'' \subset
r\Delta r'$, i. e. $r'$ is a minimal element of $Mod(\IC)$ wrt the
order $\leq_r$.

\end{document}